\documentclass[epsfig,aps,prd]{revtex4}
\usepackage{graphicx}
\usepackage{amsfonts}
\usepackage{subfigure}

\newcommand{\be}{\begin{equation}} 
\newcommand{\ee}{\end{equation}}
\newcommand{\La}{\Lambda} 
\newcommand{\T}{\tau}

\newcommand{\h}{$h^{-1}$} 
\newcommand{\lleq}{\lower0.9ex\hbox{$\buildrel < \over \sim$} ~} 
\newcommand{\ggeq}{\lower0.9ex\hbox{$\buildrel > \over \sim$} ~} 
\newcommand{\thi}{${\cal T}$}
\newcommand{\bre}{${\cal B}$} 
\newcommand{\len}{${\cal L}$}

\begin{document}
\title{Exploring the Geometry, Topology and Morphology of Large Scale Structure
  using Minkowski Functionals}
\author{Jatush V. Sheth \footnote{
This was the last paper written
by Jatush Sheth, a talented young cosmologist who met his tragic death in
a road accident on November 27, 2004.}
 \email{jvs@iucaa.ernet.in}} 
\affiliation{Inter-University Centre for Astronomy and Astrophysics,\\
  Post Bag 4, Ganeshkhind, Pune 411007, India.\\
  and \\
  Max Planck Intitute f\"{u}r Astrophysik, D-85748 Garching, Germany}
\author{Varun
  Sahni\email{varun@iucaa.ernet.in}}
\affiliation{Inter-University Centre for Astronomy and Astrophysics,\\
  Post Bag 4, Ganeshkhind, Pune 411007, India.}

\begin{abstract}
Modern redshift surveys such as the 2 degree field Galaxy Redshift
Survey (2dFGRS) and the Sloan Digital Sky Survey (SDSS) reveal the fully
3 dimensional distribution of a million or so galaxies over a large
cosmological volume. 
Visually galaxies appear to be distributed along sheet-like
and/or filamentary superclusters.  The CfA Great Wall, 
Southern Great Wall and the recently discovered SDSS Great
Wall are very spectacular superclusters. 
Clearly theoretical predictions for galaxy
clustering must be tested against these rich datasets.
This can be achieved by means of
the Minkowski Functionals (MFs).
A MF-based approach provides an excellent description of
superclusters and 
voids and allows one to quantify the properties of the cosmic web.
In this review we give a summary of the progress made in this
direction. After reviewing the status of observations and of
numerical simulations, we comment upon the nature of bias which
serves as a link between theoretical predictions and
observations. 
We also summarise the methods developed
for efficient numerical estimation of MFs for cosmological datasets
and list several important results obtained using these
methods. Specifically, we stress the discriminatory power of MFs and
of the derived morphological statistics, the Shapefinders.
Shapefinders are an excellent tool with which to study the shapes and
sizes of superclusters and voids. 
We also discuss some of the
important effects of scale-dependent bias which are brought out by
a MF-based study of the mock catalogues of galaxies. Such effects,
we note, should be accounted for before comparing theoretical models
with observations.

\end{abstract}
\maketitle
\nopagebreak
\section{Introduction}
\label{sec:intro}

During the past decade, CMB experiments such as WMAP \cite{wmap}, 
type-Ia supernova searches and observations of galaxy clustering,
have helped in turning cosmology into a
precision science. The combined set of measurements point to a {\em
  flat}, low density ($\Omega_m \simeq 0.3$), accelerating Universe dominated
by dark energy ($\Omega_{\rm DE}\simeq 0.7$). Modern redshift surveys aim
to complement the above information about background cosmological
parameters by providing a detailed map of the distribution of baryonic
matter on large scales. Redshift surveys have so far been primarily
used for two purposes: (1) to get independent constraints on the
matter density and several other cosmological parameters (see below)
and (2) to measure the abundance and clustering properties of a variety of
galaxies.  The former exercise provides important consistency tests
for cosmic parameters estimated using other methods, whereas the
latter gives precious feedback to the physics of galaxy formation. In
addition, it is also important to test whether the observed
large-scale clustering properties of galaxies are consistent with the
predictions of the preferred cosmological model(s). In order to carry
out such a test, one needs to quantify the information content hidden
in redshift space distribution of galaxies. In this article we review
the methods developed to study the geometry, connectedness and
morphology of large scale structure (LSS) in redshift surveys, which
can eventually help us meet such a goal.  Such an approach gives
valuable insights into the large-scale distribution of
matter, and provides a framework within which to test the paradigms of
gravitational instability and of the Gaussianity of the primordial density
field.

Impressive redshift surveys have been undertaken recently to map out
the 3$-$dimensional distribution of galaxies. In the next section, we
shall outline the history of redshift surveys, and focus on two major
recent surveys, 2dFGRS and SDSS.  LSS in redshift surveys exhibits
phenomenally rich and complex texture, wherein sheet-like and/or
filamentary superclusters intersperse with large, empty regions called
{\em voids}.  Several salient aspects of the LSS of the Universe will
be dwelt upon and we shall stress why it is important to study the
morphology of LSS \cite{fn1}.  In subsequent sections we shall show
that the complex morphological features of LSS are objectively
quantifiable and that morphological information supplements
traditional approach of understanding LSS using its $n-$point
correlation functions and can be employed to confront theories with
observations in an integrated manner. Statistical methods which probe
the geometry and topology of LSS include percolation analysis
\cite{sh83,shandzed89}, genus curve \cite{gott86,melotopo90}, minimal
spanning trees \cite{barbhav85}, void probability function
\cite{white79}, Voronoi tesselations \cite{rien94}, as well as more
recently introduced Minkowski Functionals (MFs)
\cite{meckwag94,Krofton97}.  For excellent reviews of these
different approaches, the reader may refer to
\cite{sc95,rien,martinez02,jones04}.  In this article, our focus will
be on MFs.

\section{Redshift Surveys}
\label{sec:surveys}

Ever since sky maps of galaxy counts, e.g., Lick Catalogue
\cite{shane54} revealed a rich pattern in LSS of galaxies in
projection, it was realised that complementing this information with
the redshifts of individual galaxies was essential to get a fully
3$-$dimensional view of the Universe.  In such an exercise, one turns
the redshift of a galaxy into its distance with the help of
Hubble's law \cite{hubble29}, and is thus able to study the {\em
  redshift space} distribution of galaxies.

Various small surveys were carried out in the late 1980's and mid 1990's in
a controlled fashion; their positive scientific outcome providing
impetus for the initiation of bigger ones. 
Of these perhaps the most influential
was the Center for Astrophysics Survey\cite{lapp86}.  This
survey had an angular coverage of $6^o\times120^o$ and a depth of
$\simeq150h^{-1}$Mpc. The survey measured redshifts of about 2400
galaxies. The famous {\em Great Wall} centered around the Coma cluster
was the most striking feature of this survey. The slice showed very
clearly the ``bubbly'' nature of LSS with voids having sharp
boundaries; the largest void having a diameter of $\simeq50h^{-1}$Mpc.
Subsequent surveys including follow-up CfA slices and the ESO Southern Sky
Survey\cite{dacosta91} amply confirmed the impression given by the first CfA
slice. Indeed, the Southern Sky Redshift Survey (SSRS) discovered a second
Great Wall in the southern sky. The emerging picture of filamentary and 
sheet-like superclusters\cite{fairall} encircling frothy voids prompted
cosmologists to seek an explanation from theory, and this interplay between
theory and observations is only now maturing into a  truly quantifiable 
discipline. Another important survey,
the Optical Redshift Survey (ORS)\cite{sant95} had
a depth of 80$h^{-1}$Mpc, but attempted a complete sky coverage
(except for the {\it zone of avoidance}). This survey measured 8500
redshifts in total, and was heavily used to estimate luminosity
functions, galaxy correlations, velocity dispersions etc.  In the mid 1990's,
the Stromlo-APM redshift survey\cite{love95} and the Durham/UKST
redshift survey\cite{rat98} were focussed towards the southern sky and
led to fruitful results on correlation functions in real and redshift
space, power spectra, redshift distortions, cosmological parameters,
bias etc. However, the galaxy redshifts in all these surveys never
exceeded $\sim 10000$.

The Las Campanas Redshift Survey (LCRS)\cite{shect96} consisted of six
1.5$^o\times80^o$ slices (3 each in the northern and southern galactic
hemisphere), went to the depth of $\simeq$750 $h^{-1}$Mpc ($z\sim0.25$)
and recorded redshifts of about 25000 galaxies. LCRS presented an
order-of-magnitude improvement over earlier surveys. This was the
first deep survey of sufficient volume which could test ideas such as the 
statistical homogeneity of the Universe on large scales. Other important
results from analysing LCRS included
the luminosity function,
second and third-order correlation functions, power spectra etc.
LCRS is perhaps the {\em only} large survey wherein both the
topology and morphology of LSS have been extensively investigated
\cite{colley97,bharad,bharad04a} and used to confront observations
with theoretical models \cite{bharad04b}. 
\subsection{The 2 degree Field Galaxy Redshift Survey (2dFGRS)}
The 2dF multi-fiber spectrograph on the 3.9m Anglo-Australian
Telescope is a very impressive facility which can obtain spectra for up to 400 objects simultaneously over
a field of view with 2$^o$ diameter. 
The spectrograph was put
to use to sample two contiguous constant-declination strips of the
sky, one in the northern galactic hemisphere [with angular coverage of
$10^o\times90^o$ in ($\delta, \alpha$)] and the other in the southern
galactic hemisphere [with angular coverage $\sim15^o\times90^o$ in
($\delta, \alpha$)].  Working in the $B_J$ band, the survey reached a
depth of $z=0.3$ with median redshift of 0.11, and has compiled
redshifts of about 2.5$\times10^5$ galaxies which are brighter than an
extinction corrected magnitude of $b_J$ = 19.45\cite{colless03}.
\begin{figure}
  \begin{center}
    \centering
    \includegraphics[width=4.0in]{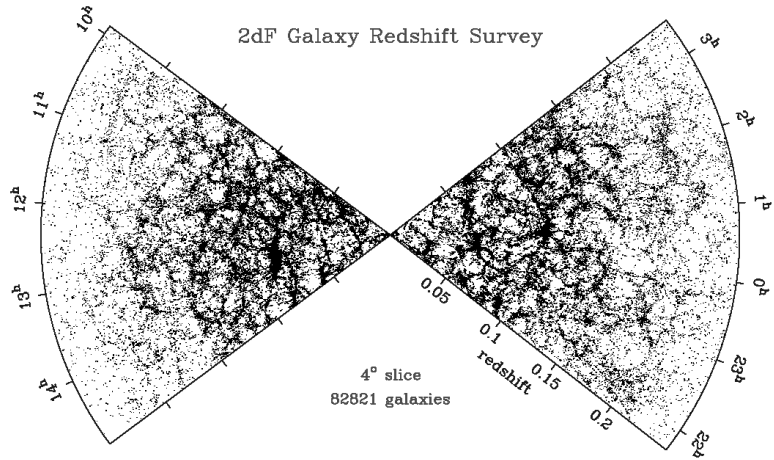}
    \caption{Flux-limited 2dFGRS slices in northern and southern 
      galactic hemisphere are shown. The supercluster-void network
      visually stands out, and can be properly understood by computing
      either a hierarchy of correlation functions or by alternate,
      geometric and topological quantifiers. In this article, we
      review methods to delineate individual superclusters and voids
      and describe their properties by using topological and
      geometrical diagnostics such as Minkowski Functionals and
      Shapefinders.}
  \label{fig:cweb}
\end{center}
\end{figure}
The survey is already complete and the resulting galaxy correlation
functions, redshift distortions and pairwise velocity dispersions
\cite{hawkins03} demonstrate the superb quality of the data set.  The
science output of this survey and its overall cosmological relevance
have been excellentlly reviewed in \cite{pea03}.  We shall discuss here
some of the results which are most relevant to our present discussion.

One of the main science goals of 2dFGRS was to get improved estimates
of the 3-dimensional power spectrum and the two-point galaxy-galaxy
correlation function\cite{fn2}. Since 2dFGRS sampled a cosmological
volume as large as $\simeq0.1$[\h~Gpc]$^3$, it became possible to
measure these quantities over a large range of scales 0.01 $\le k \le
0.4 h$ Mpc$^{-1}$ with an unprecedented
accuracy\cite{perci01,tegmark02}.  Assuming a constant bias between
dark matter and baryons on scales $k \le 0.15 h$ Mpc$^{-1}$, Percival
et al.(2001) compared the shape of the power spectrum with the
primordial power spectrum predicted by linear perturbation theory of structure
formation\cite{perci01}. This allowed them to constrain the values of
$\Omega_m$, $\Omega_B$ after priors on the power spectrum index $n$
and the Hubble constant $h$ had been applied ($\Omega_m=0.26\pm0.03,
\Omega_B=0.044\pm0.016$). Apart from this, 2dFGRS was employed to
obtain estimates for the luminosity function of early and late type
galaxies\cite{madg02}, and to study the luminosity dependent
clustering of galaxies as measured using the correlation function
statistic\cite{nor02}. The latter study provided clues about the large
scale bias between baryonic and dark matter and revealed that
the former might follow the latter through a linear relationship (also
see \cite{lahav02}).

Cosmological parameter estimation is clearly one of the key results
gleaned from large redshift surveys such as 2dFGRS. However, in our
view several assumptions made in the process remain to be adequately
justified, for instance, the assumption of linear, constant bias used
in the exercise mentioned above is clearly inadequate. In fact, the
study of biasing of galaxies with respect to the underlying mass
distribution has its own history, and the issue has not yet settled
completely. Below we shall briefly discuss the currently established
ideas about bias.

Figure \ref{fig:cweb} shows the distribution of galaxies in the
flux-limited 2dFGRS slices. To the eye, galaxies are distributed
anisotropically, with sharp qualitative similarities with the
distribution of dark matter in N$-$body simulations (see below).  The
3$-$dimensional power-spectrum, in a more appropriate form
$\Delta^2(k) = dP(k)/dln(k)$, measures the amount of power per unit
interval in the logrithmic bin around wavenumber $k$.  In simple
terms, this is a statistical measure of the density fluctuations
within spheres of radius $R\sim1/k$. Clearly, in a measure such as
this, the information on length-scales $r < R$ is {\em averaged out}.
Hence, one learns relatively little about the nature of the {\em
  actual} distribution of galaxies. Describing (and explaining) the
visually rich filamentary pattern of superclusters of galaxies and
quantifying the complex nature of the LSS is another, {\it equally}
important application of redshift surveys.  This problem has not been
adequately addressed despite considerable recent progress
\cite{hoyle04, hoyle02, hoyleetal02, hikage03a, basil03, dorosh04,
  hikage02}.  Clearly, the contribution of higher order correlation
functions is an important ingredient in such a program. Since higher
order correlations are difficult to measure in practice, it would be
useful if we could quantify the connectedness and non-Gaussianity of
LSS as well as the shapes and sizes of individual superclusters and
voids using geometrical diagnostics.  The methods reviewed in the
present article are intended to fill this gap in our knowledge and to
complement more traditional, correlation function based approaches in
developing a quantifiable picture of the LSS and its properties.
\subsection{Sloan Digital Sky Survey (SDSS)}
The Sloan Digital Sky Survey (SDSS) is derived from a dedicated 2.5m
telescope.  The initial photometric program has been to measure the
positions and brightness of about 10$^8$ objects in $\pi$ steradian
(almost 1/4$^{\rm th}$) of the sky. The survey is centered on the
northern galactic pole and has an elliptical angular coverage of
$130^o\times110^o$, where the semi-major axis runs along a line of
constant right-ascension.

Follow-up spectroscopy is planned to give redshifts of about 10$^6$
galaxies and 10$^5$ quasars.  At the time of writing, the SDSS team
has made its 2$^{\rm nd}$ data release, available for free download
\cite{abaz04}.  It contains redshifts of $\simeq3\times10^5$ galaxies,
with the farthest galaxy at a redshift of $\sim$0.3. (See
\texttt{http://cas.sdss.org/astro/en/tools/search/SQS.asp}.)

The limiting magnitude of the SDSS survey is in the red band ($r_{lim}
= 17.77$), and its depth is similar to 2dFGRS. The survey will cover a
cosmological volume as large as 1[$h^{-1}$ Gpc]$^3$. Taking the length
scale of 100 $h^{-1}$Mpc as a tentative scale of homogeneity, this is
approximately thousand times a representative volume of the Universe.
Hence when fully complete, SDSS can be fruitfully used to evaluate
the cosmic variance for many statistics associated with LSS. The data are
already large enough to permit accurate estimation of power spectrum
\cite{tegmark04a}, 3$-$point correlation function \cite{kayo04}, and
of cosmological parameters \cite{tegmark04b}. Complementary studies of
LSS in SDSS have been undertaken using MFs by \cite{hikage03a} and
using the minimal spanning tree-formalism by \cite{dorosh04}. We shall
mention later some results from an MF-based study of LSS in {\em mock}
catalogues of SDSS \cite{jsh04}.

In order to make full use of the information encoded in the observed
LSS, it is essential to understand the relationship between the number
density of galaxies and the underlying mass density field. Since
1980s, it has been realized that the {\em visible} density field due
to galaxies may be {\em biased} relative to the mass-density.  In
order to appreciate the implications that bias may have in the program
of confronting theory with observations, below we briefly review the
state of our knowledge in this regard. First we shall discuss the
theoretical modelling of LSS using numerical simulations.

\section{N$-$Body Simulations}
\label{sec:sec3}

Strong evidence exists to suggest that dark matter is $\sim$10 times more
abundant than visible baryonic matter. Dark matter is usually assumed to be
collisionless and to interact only through gravitation.  Due to its relative
abundance dark matter is clearly more important than baryons from a
dynamical point of view. The best
way to understand the evolution of a gravitating system consisting of
N dark matter particles would be to {\em actually} solve for the force
acting on a given particle due to the remaining particles.  In a
rather different form, by solving for the Poisson's equation at every
stage of evolution of the matter-field, this is precisely the goal
which modern versions of N$-$body simulation codes achieve. (For an
excellent introduction to N$-$body simulations, see \cite{bert98,
  kly00a, kly00b, kly00c}).

Simulations allow us to probe the strictly nonlinear regime of the
dark matter density field, and can provide us with detailed
predictions of the {\em dark side} of the Universe at today's epoch.
(The resulting dark matter distribution can be probed through observations of 
gravitational lensing -- an area of great future promise.)
The large scale structure distribution in dark matter could very well 
\begin{figure}
  \begin{center}
    \centering \includegraphics[width=4.0in]{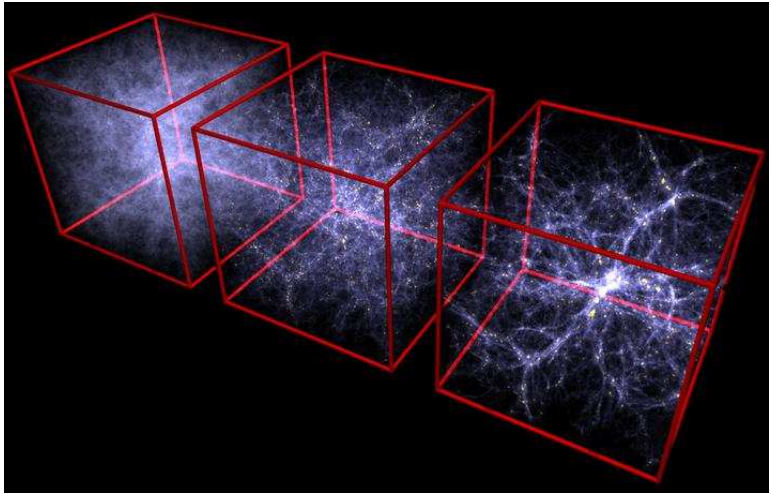}
    \caption{The figure illustrates the development of nonlinear
      structures in a hydrodynamic N$-$body simulation from $z=6$
      (leftmost cube) to $z=0$ (rightmost cube) via $z=2$ (middle
      cube). A near featureless density field evolves to produce
      filamentary and sheet-like superclusters which percolate through
      the box-volume. These are separated by large voids. Due to its
      web-like appearance, LSS is dubbed as the {\em Cosmic Web}.
      [Figure courtesy: Volker Springel; see {\it
        http://www.mpa-garching.mpg.de/galform/data\_vis/index.shtml}]}
    \label{fig:web} 
  \end{center}
\end{figure}
resemble that shown 
in Figure \ref{fig:web}. We note from this figure, how the matter
condenses along filamentary/sheet-like ridges between clusters and how
voids get progressively more evacuated with the passage of time.  
Due to web-like and 
frothy appearance, LSS has been described at various times as being a
{\it Cosmic Web}~\cite{bond96} or as {\it Cosmic Froth}~\cite{rien}.

During the past two decades the advent of parallel supercomputers have
ensured a phenomenal improvement in the computing power available to
tackle complex astrophysical processes. Hence, it has become possible
to include complex gas-dynamical processes into simulations.  Various
recipes for cooling, star formation and feedback due to supernovae and
galactic winds have been incorporated to mimic the formation and
evolution of galaxies in a cosmological setup with simulated volumes
ranging from 50 to 100 $h^{-1}$Mpc \cite{wein02,wein04}. This has made
it possible to develop {\em mock} catalogues of galaxies inspired by a
given cosmological model.  Relatively less time-consuming semianalytic
methods have also been proposed which hope to achieve similar goals
while bypassing the complex baryonic physics
\cite{benson02,cole00,kauf97,sc95,scocci02}. Such mock catalogues can
eventually be compared with clustering properties of galaxies derived
from redshift surveys with the help of methods reviewed here.

\section{The Nature of Bias}
\label{sec:sec4}

Above we summarised the theoretical efforts being made to understand
the clustering of dark matter. From the predictive point of view,
given the fully evolved nonlinear mass-density field
(both in the real universe and in N-body simulations), we would like to
know the exact locations where galaxies are formed. An important
question related to this is whether the clustering properties of the
resulting galaxy distribution (quantified in terms of $n-$point
correlation functions or some other alternate statistics) are {\em the
  same} as the underlying matter-field, or whether the two are different. The
answer to this, clearly depends upon the recipe which we utilise in
grafting galaxies on the density field.

During the mid-1990's it was shown that AGNs, IRAS-selected galaxies and galaxies
selected in the optical bands cluster differently \cite{pea94}.
Earlier (during the 1980's) it was established that clusters of
galaxies are more strongly
clustered than galaxies themselves \cite{kai84}. Recently it was proven that the
more luminous galaxies are more strongly clustered \cite{nor01,
  nor02}. Since these different visible tracers of LSS appear to cluster
differently, their distributions are clearly biased relative to each
other. There is hence, a reason to believe that the visible matter as
a whole will be biased relative to the underlying dark matter. This
phenomenon is termed as {\em bias}.  Knowledge of the nature of
bias will clearly enable us to connect the two distributions $-$ visible and
dark; observed and simulated.

In cosmology, it is a conventional practice to quantify the clustering
of LSS using the two-point correlation function or, equivalently, the
power spectrum. Both these
statistics are known to have a power-law form for galaxies
\cite{groth77,baugh96}. However, the correlation function of dark
matter was shown to differ from $\xi_{gg}(r)$ in a complicated,
scale-dependent fashion for various
CDM-based cosmogonies\cite{kly96,pea97,jenkcdm}. This suggests that the bias
{\em cannot be} scale-independent, i.e., light may not 
follow the mass in a linear fashion. 
Recent investigation involving the three$-$point correlation evaluated for the
SDSS dataset implies that galaxy biasing could be quite
a complex and nonlinear process; see
\cite{kayo04}.

Kaiser(1984) explained the strong clustering of clusters using the
{\em high peak model} \cite{kai84}. This model states that a rare,
high-density fluctuation corresponding to a massive object of a given
size, collapses sooner if it lies in a region of a larger-scale
overdensity.  This is because, the small-scale overdensity is {\em
  aided} by the surrounding large-scale overdensity so that the
collapse-criteria is fulfilled relatively easily for such a halo. This
``helping hand'' from the long-wavelength mode means that overdense
regions contain an enhanced abundance of massive objects with respect
to the mean. Later this model was generalised to incorporate objects of
any mass \cite{cole89, mowhite96}. 

The high-peak model is based on kinematic premises. It traces back the
observed clustering of various classes of objects to the in-built
structure in the primordial density field. It could fruitfully be
applied to study the clustering of galaxies only at high redshift
($z\simeq2.5-3$) \cite{kai84,cole89,steidel97}, for the galaxy-scale
fluctuations were prone to collapse at such early times. At such high
redshifts, galaxies were shown to be highly biased relative to
background \cite{steidel97}.  Starting from $z\ggeq3$, until the present,
galaxies have very likely collapsed in all environments. The underlying
mass-fluctuations have also grown. Thus, one predicts that the large
bias (defined as $b(r) = \sqrt{\xi_g(r)/\xi_{dm}(r)}$) which existed
earlier, should asymptotically approach unity on large scales.  Lahav
et al.(2002) showed for 2dFGRS data that the bias averaged over all
scales is statistically consistent with a scale-independent,
large-scale bias of order unity. However, we should see whether this
knowledge helps us achieve our original goal of connecting the
simulations with the observed Universe, i.e., in populating the
mass-density field with galaxies in the correct manner.

A recent, {\em Halo Model} based approach is a fruitful step in this
direction. Here all the complications of galaxy formation are encoded
via the halo occupation number: the number of galaxies found above
some luminosity threshold in a virialised halo of a given mass . In a
nut-shell, the halo model describes nonlinear structures as virialized
halos of different mass, placing them in space according to the
linear, large-scale density field \cite{cooray02}. This model is
analytically tractable, and correctly incorporates the dependence of
bias on the 2$^{\rm nd}$ order correlations in the density field.
Consequently, it correctly reproduces the observed 2$-$point
correlation function and the power spectrum.
\begin{figure}[h]
  \begin{center}
    \centering
    \includegraphics[width=4.0in]{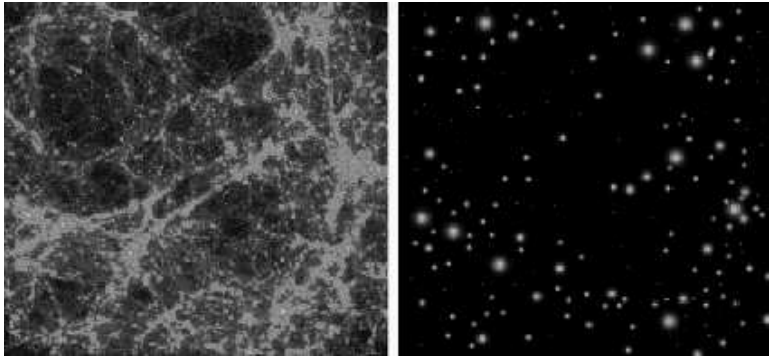}
    \caption{{\it Left panel}: Simulated dark matter distribution which
      shows complex morphology. {\it Right panel}: Halo Model
      corresponding to density field in the left panel, which
      successfully reproduces galaxy-galaxy correlation function and
      the power spectrum. The two are manifestly different, which
      points out the role which higher order correlation functions
      play in defining the cosmic structure. Figure courtesy \cite{cooray02}.}
    \label{fig:halomodel}
  \end{center}
\end{figure}
Figure \ref{fig:halomodel} compares the exact non-linear mass-density
field with the halo model representation. The halo model is remarkably
successful in obtaining correct correlation function, power spectrum
etc. However, as shown in Figure \ref{fig:halomodel} it is manifestly
different from the fully evolved density field. This brings out an
important aspect relevant to further study of LSS: the low-order
correlations cannot be relied upon to provide us with a complete
picture of clustering. In addition to the large halos of clusters and
groups, today the rich filamentary/sheet like patterns in matter
density field are also populated by galaxies\cite{fn4}, and one must
theoretically understand the emergence of halos along filaments/sheets
in order to complete the picture. In fact, the following experiment
can be carried out: the galaxies can be {\it selected} in $z=0$
mass-density fields from two rival cosmogonies, so that the observed
galaxy-galaxy correlation function $\xi_{gg}(r)$ and the power
spectrum $P_k$ are reproduced.  In comparing our predictions with the
LSS-data, if we confine ourselves merely to two-point correlation
function, such an algorithm would validate {\em both} models to be
good representatives of our Universe. Evidently this is not true.
Clustering properties of galaxies in the two models would of course be
different from each other (and perhaps even from LSS in the Universe).
This serves to establish that the large-scale distribution of galaxies
{\em cannot} solely be described by the two-point correlation function
alone \cite{fn6}.  The resulting galaxy distributions can be shown to
differ from each other by employing statistics which are sensitive to
higher order correlation functions. Following this, we arrive at an
important conclusion: the two distributions of galaxies {\em must be}
compared with the help of higher order correlation functions or
statistics derived therefrom.

To summarise, although the {\em high peak model} \cite{kai84} gives us
promising results for galaxies assembling at high redshift and for
Abell clusters at present, the predictive power of {\em halo model} in
populating mass density field with galaxies is clearly inadequate (in
fact, due to a variety of reasons, the bias could even be stochastic;
see \cite{dekel99}). In particular, for a combination of a
cosmological model and a biasing scheme to be proven correct, it is
{\em not sufficient} that the two explain the two-point correlation
function for galaxies {\it alone}. In order to effectively compare the
theoretical predictions for galaxy-distribution with the data from
redshift surveys, one should include information from higher order correlation
functions, and/or alternate statistics which can effectively quantify the
complex morphology of the galaxy distribution. One such class of diagnostics
the Minkowski Functionals -- are introduced  in the next section.

\section{Minkowski Functionals}
\label{sec:sec5}

\subsection{Minkowski Functionals in Cosmology}
It is well known that, in contrast to a Gaussian random field
(hereafter, GRF), a fully evolved nonlinear density field cannot be
fully quantified in terms of its two-point correlation function; the
latter being simply the lowest and first of an infinite hierarchy of
correlation functions describing the galaxy distribution. Furthermore,
the bias in galaxies viz a viz dark matter can be nontrivial in
nature, and can give rise to different clustering properties of
galaxies and the underlying mass density field. Consequently the
correlation function $\xi_{gg}(r)$ determined either for galaxies or
for dark matter does not on its own validate a given cosmological
model unless support is provided by other, statistically indepdend
measures of clustering. It therefore becomes a challenge to compare
the theoretical predictions for LSS with real observational data.

If we knew all $n-$point correlation functions, we would have a
complete description of the galaxy clustering process. However,
estimating $\xi_{gg}(r)$ for a sample of $N$ galaxies requires knowing
all pairs of galaxies in this sample, whereas calculating the 3-point
function implies taking all triplets. The amount of computation
escalates rapidly with $n$ -- the order in the $n-$point correlation
function -- especially for a large value of $N$. Besides, it is difficult
to extract intuitively useful information from these statistics. The
present article focuses on another class of statistics which
complement the correlation function approach and which have the
advantage of providing a physically appealing interpretation for the
evolving density field.  These are the {\em Minkowski Functionals}.

For an excursion set involving particles embedded in
$n-$dimensions, the Minkowski Functionals (hereafter, MFs) are defined
on an ($n-1$)-dimensional hypersurface. There are $n+1$ MFs in $n$ dimensions. In this
article, we shall be concerned with the 3-dimensional distribution of dark
matter and/or galaxies. Hence, MFs will be defined on a 2-dimensional
surface and they will reflect the physical properties of this surface
(in the given instance, an isodensity contour referring to a
supercluster of galaxies or a void).  For a given surface the four MFs \cite{blaschke36,
  min1903, meckwag94} are, respectively
\begin{enumerate}
\item Volume V,
\item Surface area S, 
\item Integrated Mean Curvature C,  
  \begin{equation}
    C = {1\over2}\oint\left({1\over R_1} + {1\over R_2}\right)dS.
  \end{equation}
  $R_1$ and $R_2$ are the two principal radii of curvature of the
  surface in a given local neighbourhood and the integral is taken
  over the entire closed surface.
\item Integrated Gaussian Curvature (or Euler characteristic) $\chi$, 
  \begin{equation}
    \chi = {1\over2\pi}\oint\left({1\over R_1R_2}\right)dS.
  \end{equation}
  A related quantity which is more popular in Cosmology is the genus $G$.  The
  genus is related to $\chi$ by
  \begin{equation}
    G = 1 - {\chi\over2}.
  \end{equation}
  The 3-dimensional genus of an object is a topological invariant.  It
  can be interpreted in terms of the connectivity of the surface.  In
  simple terms it can be viewed as the number of independent cuts
  which one can make to the surface {\em without} breaking it into two
  separate pieces. A torus can be cut {\em once} and yet remain in one
  piece. Hence, its genus is 1. However, a sphere would separate into
  two pieces if a cut were made. Hence, its genus is 0. Two objects with
  the same value of genus are topologically similar: one can be
  obtained by continuously deforming the other. Thus, a sphere and
  a cube are topologically equivalent.
  More concretely, genus of an object is the number of handles that
  the object has, in excess of the number of holes which it encloses
  (e.g., see \cite{matsub03}). Thus,
  $$
  {\rm G = [\# ~of ~handles ~to ~the ~surface] - [\# ~of ~holes ~enclosed ~by ~the ~surface].}
  $$
  According to this definition, a sphere has no handle, a torus has
  a single handle (equivalent to a sphere with one handle), and a
  pretzel has two handles. Introducing a hole or a bubble {\em inside}
  the surface {\em reduces} its genus by 1, whereas adding a handle to
  the surface {\em increases} its genus by 1.
\end{enumerate}
  
An attractive feature of MFs is that they depend upon
the entire hierarchy of correlation functions \cite{schmalthesis}.
Thus, MFs can indirectly help us capture the effect of $n-$point
functions.  Higher order $n-$point functions gradually become
important as a primordial, featureless GRF evolves to develop
nonlinear structures.  This reflects in the behaviour of
MFs for the corresponding density field. MFs are additive in nature,
i.e., they can be studied for individual objects (say, clusters or
superclusters defined using some prescription) as well as for the
entire ensemble of such objects.

Of the 4 MFs listed above, the genus G was already known to the
cosmology community \cite{dorosh70, gott86, melotopo90}.  Tools like
percolation analysis which can also be related to MFs, had earlier
been introduced in cosmology by \cite{zes82}. There were strong
reasons why this was the case.

It is well known that a system evolving under gravitational
instability becomes progressively more non-Gaussian. (The hypothesis
of a primordial spectrum of density perturbations which is distributed
in the manner of a Gaussian random field is supported by observations
of the cosmic microwave background carried out by WMAP and other
experiments.)  For CDM-like cosmological models this results in more
matter being concentrated in filamentary and pancake-like
distributions whose coherence scale evolves with time.  As a result
the filling fraction at percolation progressively decreases, from
$16\%$ for the initial Gaussian random field to $\sim 5\%$ for LCDM
\cite{ks93,sss97,shyes98,sh-sh-s04}.  This implies that more matter is
being transferred to regions which occupy a progressively smaller
amount of space and are also spatially anisotropic and therefore
percolate easily.  Figuratively this corresponds to the prominance
acquired by superclusters of galaxies over time -- see Figure
\ref{fig:web}.  This theoretical picture which described the emergence
of structure in an initially featureless medium agreed well with what
observations were telling us about our universe.  The emerging picture
of LSS taken from redshift surveys shows us that galaxies are
distributed preferentially along filaments and sheets which seem to
encircle vast voids. The distribution of galaxies does not occupy much
volume, and because of the presence of superclusters, percolates
easily.  Percolation theory can therefore be used to describe an
important quality of the observed galaxy distribution -- its
connectivity.  At the percolation threshold one can visit the far
corners of the cosmic web by following filamentary/sheet-like
overdensities. The cosmic web is therefore a connected structure and
this is clearly revealed in surveys such as Figure \ref{fig:web}.
Connectivity is of course a mathematical notion, and it is possible to
precisely quantify connectivity of an object or a system by evaluating
its percolation properties and its genus.  Doroshkevich(1970) gave
analytic formula for the genus of a GRF which showed that at the
median density threshold, GRF would exhibit a sponge-like topology
\cite{dorosh70}. Simulations of LSS within CDM-cosmogonies provided
similar visual impression when smoothed on sufficiently large scales.
Like percolation, the genus too has come to be considered as a useful
probe of the initial Gaussianity (as well as the final
non-Gaussianity) of the cosmic density field; see for instance
\cite{melotopo90, colley97,sss97}.

\begin{figure}
  \begin{center}
    \centering
    \includegraphics[width=4.2in]{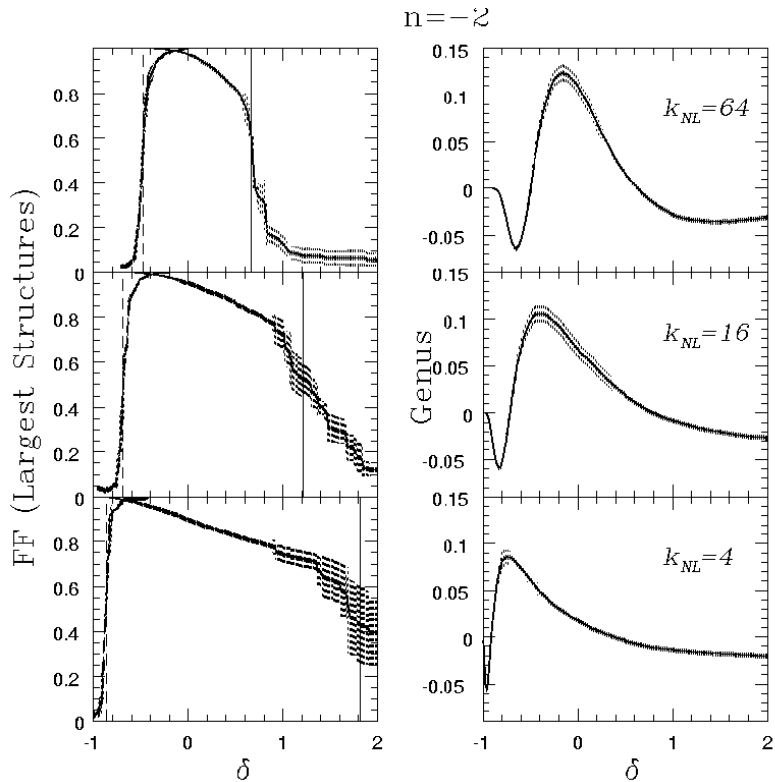}
    \caption{Percolation (left panels) and genus (right panels) curves
      are plotted as functions of the density contrast $\delta$ for
      scale free models of gravitational clustering with the
      perturbation spectral index $n=-2$. Solid and dashed curves in
      the left panels correspond to the percolation curve for the
      largest cluster and void respectively. Vertical solid/dashed
      lines mark the threshold describing percolation between opposite
      faces of the cube for clusters/voids respectively. Figure courtesy of
      \cite{sss97}.}
    \label{fig:perc_gen}
  \end{center}
\end{figure}
An important indicator of non-Gaussianity in a distribution is the
{\em percolation curve}. The percolation curve describes the {\it
  volume fraction} (or filling factor -- FF) in the largest structure
(cluster/void) as a function of the density contrast threshold
$\delta$ \cite{fn7}.  Salient features of the percolation curve are
illustrated in Fig. \ref {fig:perc_gen} in which density fields (with
an initial power spectrum $P(k)\sim k^{-2}$) evolve from an epoch when
the scale of nonlinearity is $k_{\rm NL}= 64k_f, 16k_f, 4k_f$, ($k_f$
is the fundamental mode corresponding to the box-size of our N-body
simulation).  In Fig. \ref {fig:perc_gen} percolation curves for
clusters (thick solid lines) and voids (thick dashed lines) are shown
as functions of the density contrast $\delta.$ Starting from a high
density threshold (small $FF$) we find several isolated clusters
(corresponding to peaks of the density field).  Lowering the density
threshold further results in the {\it merger} of clusters leading to a
rapid growth in the percolation curve and to the onset of percolation.
A further lowering of the threshold to very small values results in
the merger of almost all clusters so that $FF \rightarrow 1$.  An
identical procedure followed for underdense regions by gradually
increasing the density contrast threshold (increasing $FF$) results in
a similar functional form for the volume fraction in the largest void.
In our samples the largest cluster percolates between opposite faces
of the cube when its filling factor is about half the total $FF$.  In
most cases percolation also coincides with the highest jump in the
volume of the largest cluster which was used by \cite{delgh91} as a
working definition of the percolation threshold.

The solid/dashed vertical line in Fig. \ref {fig:perc_gen}
represents the density contrast threshold $\delta_C$ below/above which
clusters/voids percolate.  We note that {\it both} clusters and voids
percolate over a range of overlapping density contrasts --- a feature
that is only possible in three or more dimensions --- and corresponds
to what is commonly called a {\it sponge} topology for the density
distribution.

As the simulation evolves 
$\delta_C$ increases monotonically
for $n=-2$, as structures form and align on increasingly larger scales.
From Fig. \ref {fig:perc_gen} we see that voids
find it easier to percolate as the simulation evolves, as a result
the range in densities when both phases percolate initially increases,
enhancing the extent of {\it sponge-like topology} in the distribution.

The right panel in Fig. \ref {fig:perc_gen} shows the evolution of the
genus curve for the same simulation. The top panel corresponding to an
early epoch shows the genus curve largely retaining its bell-shaped
form which it has for the primordial Gaussian random field $G(\nu) = A
(1 - \nu^2) \exp(-\nu^2/2)$ (\cite{hgw86,gwm87}).  During later epochs
the density distribution grows progressively more non-Gaussian and
this is reflected in the change in both shape as well as amplitude of
the genus curve in Fig. \ref {fig:perc_gen}.

Although a study of the genus gives us a useful handle on the
connectivity of a density field, we still lack information about the
morphology of LSS.  For example, if the domain of information is
solely restricted to knowing genus, a filament with one handle would
be considered to be identical to a sphere with one handle or a pancake
with one handle.  Therefore, to gain objective insights into the
nature of the supercluster-void network, we must complement genus of
an object with quantities which in some way, characterise geometry of
that object.  This is precisely the role played by the first three
MFs, Volume V, Surface Area S and Integrated Mean Curvature C. These
MFs change when local deformations are applied to the surface, and
hence, these may be useful to glean information about the typical size
of an object.  The final goal here would be to use this geometric
information and give an objective meaning to generic sizes and shapes
of superclusters and voids belonging to LSS.  In this article we will
show that MFs can be utilised to characterise the geometry and
topology of the cosmic density field, and eventually can be employed
to get information about the morphology of the supercluster-void
network in LSS. Because MFs depend on the full hierarchy of
correlation functions, we can use them to compare and distinguish two
rival cosmological models. Recently, Matsubara (2003) derived
semianalytical expressions of MFs for weakly nonlinear cosmic density
field by using a second-order perturbative formalism \cite{matsub03}.
The results obtained by Matsubara can be used in conjunction with the
methods reviewed here to obtain insight into the dynamics of the
clustering process of a density field smoothed on sufficiently large
scales. The role of bias on large scales can also be directly probed
using such methods.

\subsection{Morphology of LSS with Shapefinders}

Sahni, Sathyaprakash \& Shandarin (1998) showed that the
MFs could be used to evaluate both the size as well as the shape of a 
three-dimensional object such as a supercluster or a void \cite{sss98}. 
The size is given in terms of
three Shapefinders which have dimensions of length. These are defined
as ratios of MFs, and are conviniently termed as Length (${\cal L}$),
Breadth (${\cal B}$) and Thickness (${\cal T}$).
\begin{eqnarray}
{\cal T} & = & {3\times V\over S}\\
{\cal B} & = & {S\over C} \\
{\cal L} & = & {C\over 4\pi G}.
\end{eqnarray}

The three Shapefinders are defined using spherical normalization, so
that for a sphere of radius $R$, ~${\cal T} = {\cal B} = {\cal L} = R$.

The above measures quantify the {\em size} of the object in question. To
further quantify the {\em shape} of these objects, two dimensionless
Shapefinders have been defined as follows. These are Planarity (${\cal
  P}$) and Filamentarity (${\cal F}$).
\begin{eqnarray}
  {\cal P} &=& {\cal {B - T\over B + T}}\\
  {\cal F} &=& {\cal {L - B\over L + B}}.
\end{eqnarray}

Sahni et al.(1998) showed that (${\cal P,F}$) quantify the shape of
both simple as well as complicated objects. Thus, for an oblate ellipsoid its
planarity is relatively large and its filamentarity is small.  For a
prolate spheroid, the reverse is true. The statistics respond
monotonically to deformations of these surfaces.  

If we include the genus then, the triplet of numbers (${\cal P,F,G}$), 
can define a three-dimensional
{\em Shape-Space} which may be used to represent the distribution of
shapes of superclusters and their topologies in a given cosmological
density field.
Note that the inclusion of genus is significant, since most
superclusters at moderate density thresholds can have a spongy texture
with several loops and/or branches emanating from a central body (see
Figure \ref{fig:perclcdm} for an illustration).
\subsection{Methods of Estimation}
In cosmology the distribution of galaxies in space can frequently be regarded as
being an example of a point process.
There have been several attempts of evaluating MFs for a given point
process. For example, the Boolean grain model \cite{meckwag94,
  schmalthesis} would decorate the input set of points with spheres of
varying radii, and would study the morphology of the structures which
contain overlapping spheres. According to another approach 
\cite{Krofton97} one smooths the distribution of points
using a suitable window function, and defines the density field on a
grid. MFs can be evaluated for connected structures defined on a grid
by using Crofton's formulae or Koenderink Invariants \cite{Krofton97}.

A {\em new} approach for determining MFs has been reported in
\cite{sh-s-s-sh03}.  Rather than working with a grid-approximation for
connected structures, these workers go a step forward, and model
surfaces for these objects. Minkowski Functionals are evaluated for
the isodensity surfaces {\em online}, i.e., while these are being
modelled using an elaborate surface triangulation technique. Sheth et
al.(2003) describe the algorithms for surface construction and for the
evaluation of MFs on these surfaces.  Sheth (2004) discusses the
method of implementation of these algorithms which has culminated into
a robust and accurate software SURFGEN (short for ``SURFace
GENerator'')\cite{sheth04}. SURFGEN has been tested for accuracy
against Gaussian random fields and against a variety of simply and
multiply connected eikonal surfaces. SURFGEN evaluates the geometric
MFs to better than 1\% accuracy in most cases, while the genus is
evaluated exactly \cite{sh-s-s-sh03}. SURFGEN has since been employed
to study geometry, topology and morphology of LSS {\em both} in dark
matter and galaxies within a variety of cosmogonies.
(SURFGEN could also be relevant in areas of science where the properties of
surfaces becomes important such as in medical imaging (tomography) and in 
condensed matter physics.)

In the next section we shall discuss a set of new results obtained by
applying MF-based techniques to LSS-data.

\section{Results}
\label{sec:sec6}
\subsection{Morphology of Large Scale Structure using Shapefinders}
Given a cosmic density field, Minkowski Functionals (MFs) can be
evaluated for a supercluster (void) defined as a connected overdense
(underdense) region above (below) certain physically motivated
threshold of density.  SURFGEN-like technique would isolate such a
structure, and would model its isodensity contour in order to
precisely estimate its MFs. MFs referring to a given structure are
called {\em partial MFs}. By knowing partial MFs, one can further
measure shape and size of the concerned structure using Shapefinders.
Further, MFs are additive in nature. Hence, at any given threshold of
density, {\em global MFs} pertaining to the entire density field can
be obtained by summing over {\em partial MFs} of the constituent
overdense/underdense objects.  While partial MFs are useful to measure
shapes and sizes of individual superclusters and voids, global MFs can
be employed (1) to discriminate between models, (2) to probe the
effect of nonlinear gravitational evolution on the density field and
(3) to capture the nontrivial effects of bias on the structure of the
cosmic web. We shall review here some of the promising results
obtained in these directions.
\subsection{Discriminating between rival models of LSS}
The growth-rate of density perturbations differs between various
cosmogonies. As a result, the clustering properties of the
matter-distribution also turn out to be different. A serius candidate
statistic of LSS should be able to capture such differences and
discriminate between rival cosmological models.

\begin{figure}
  \begin{minipage}[t]{.99\linewidth} \centering
    \includegraphics[width=5.2cm]{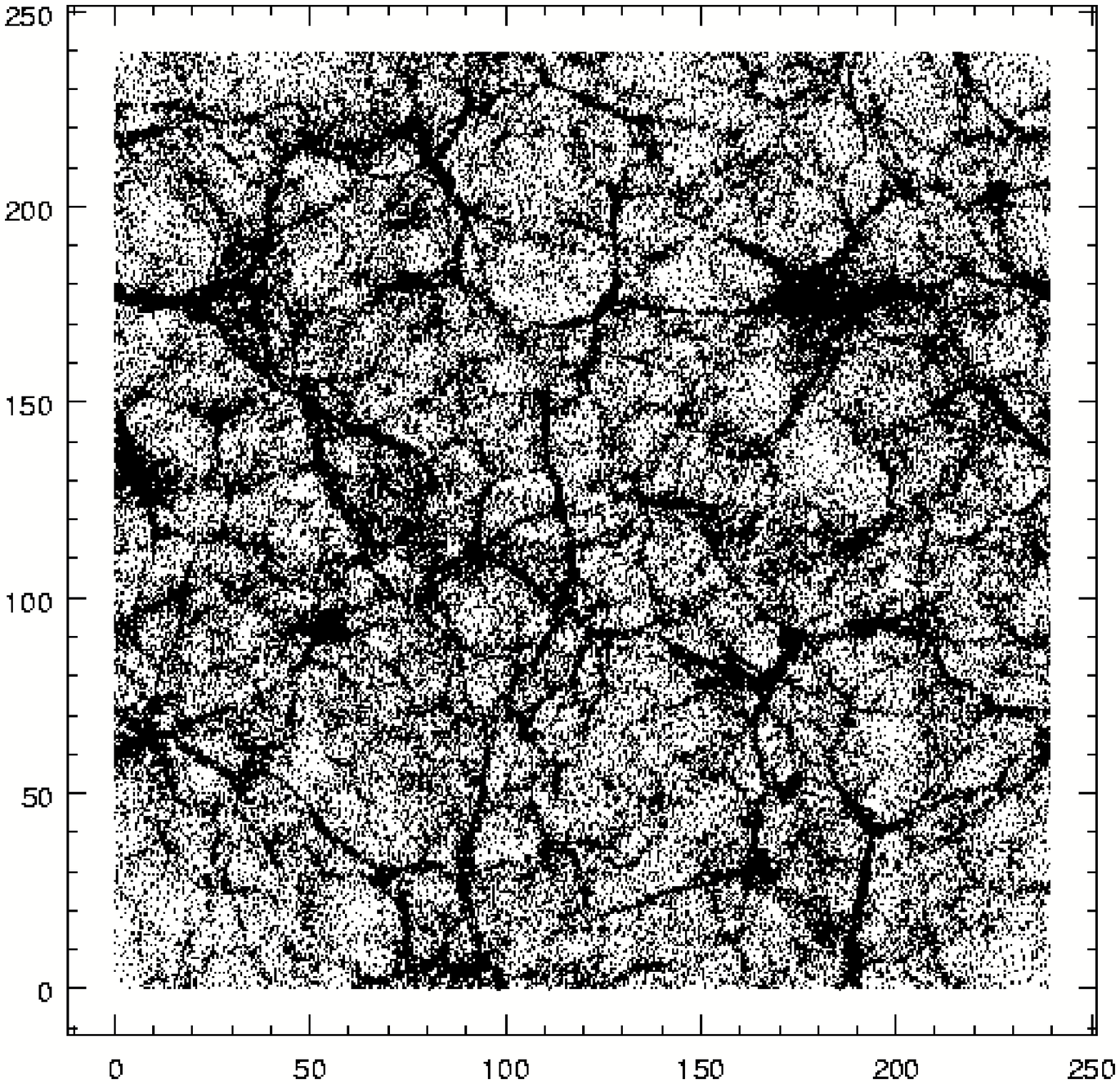}
    \includegraphics[width=5.2cm]{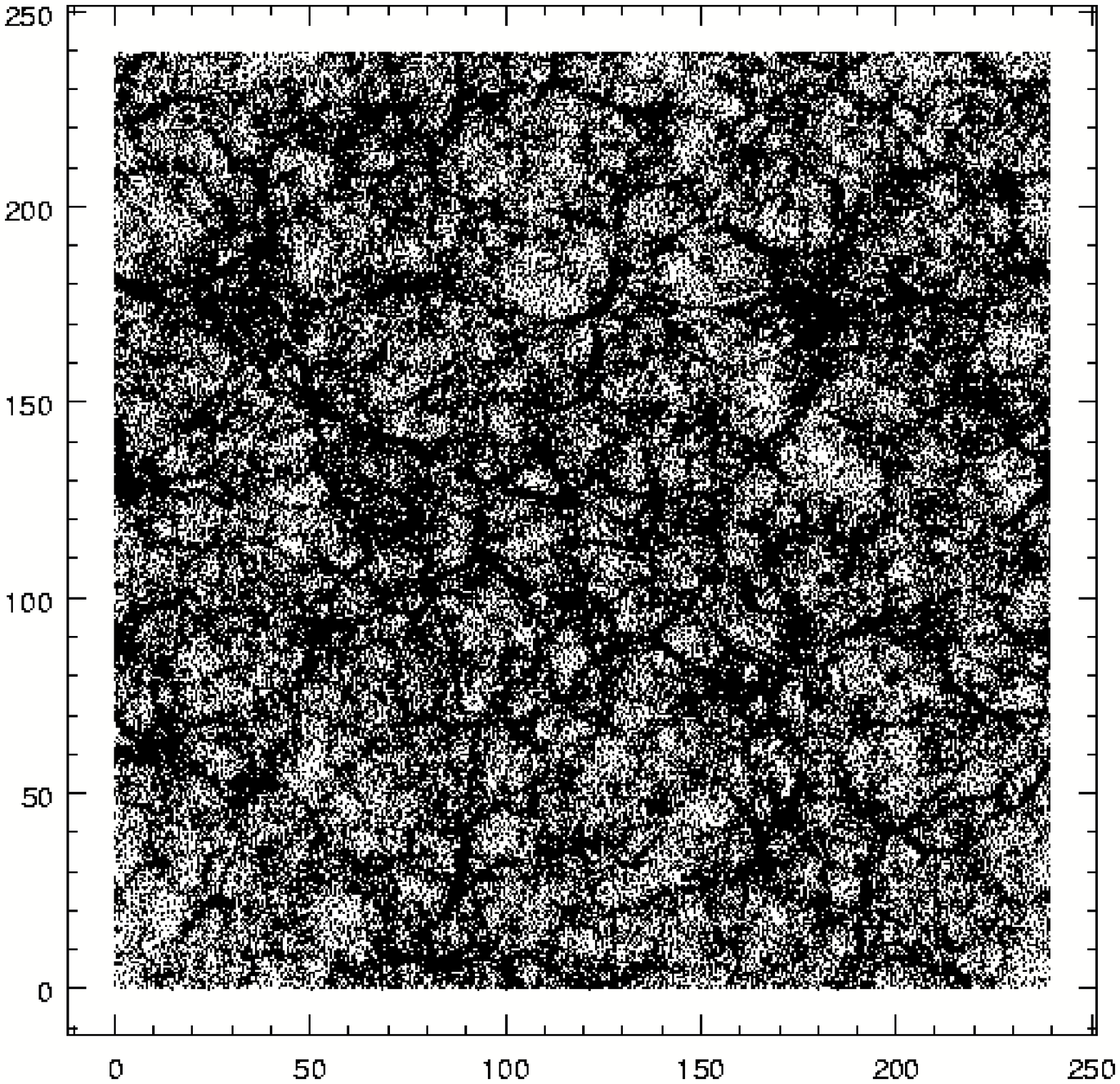}
    \includegraphics[width=5.2cm]{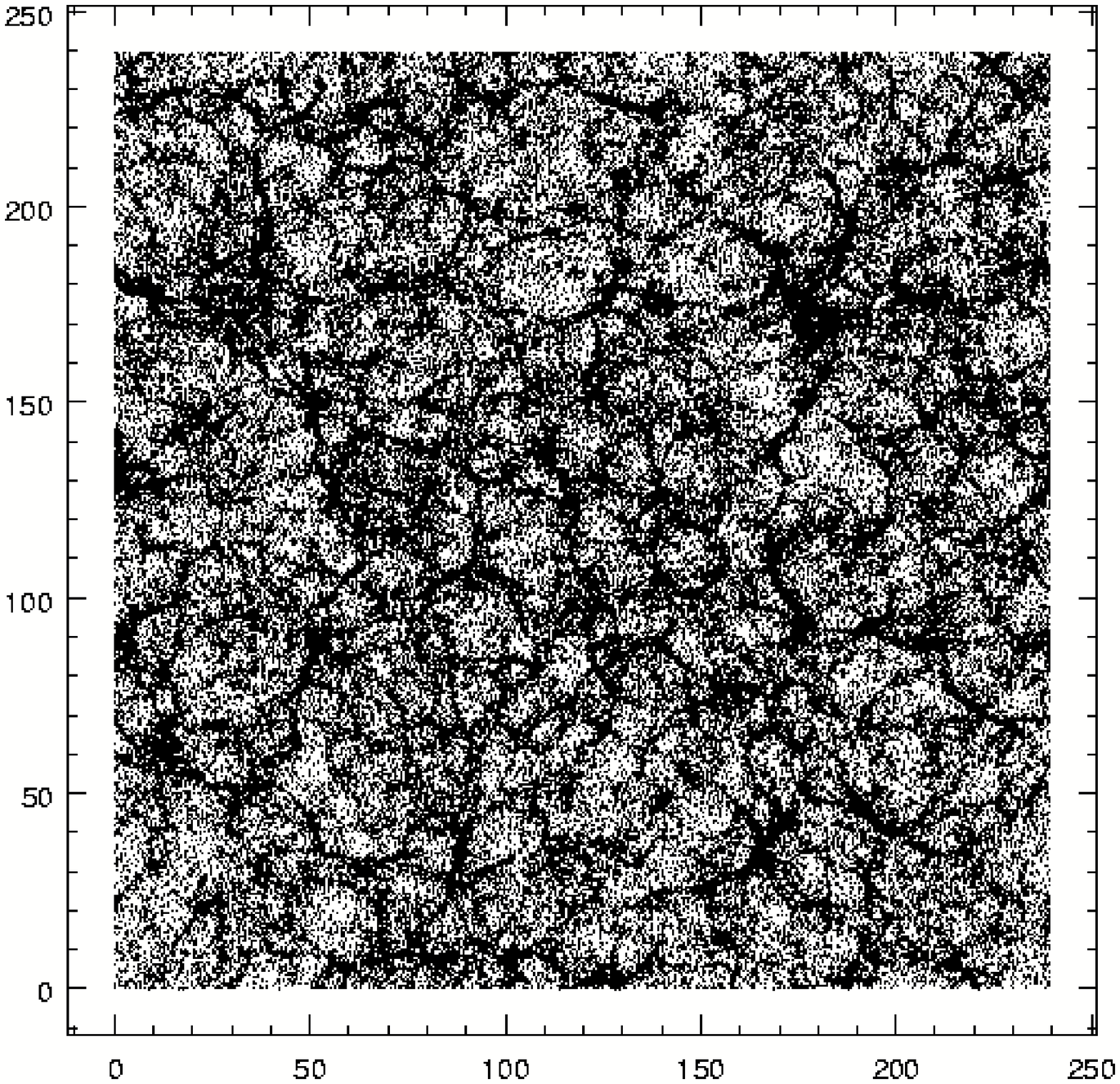}
  \end{minipage}\hfill
  \begin{center}
    \includegraphics[width=5.2in]{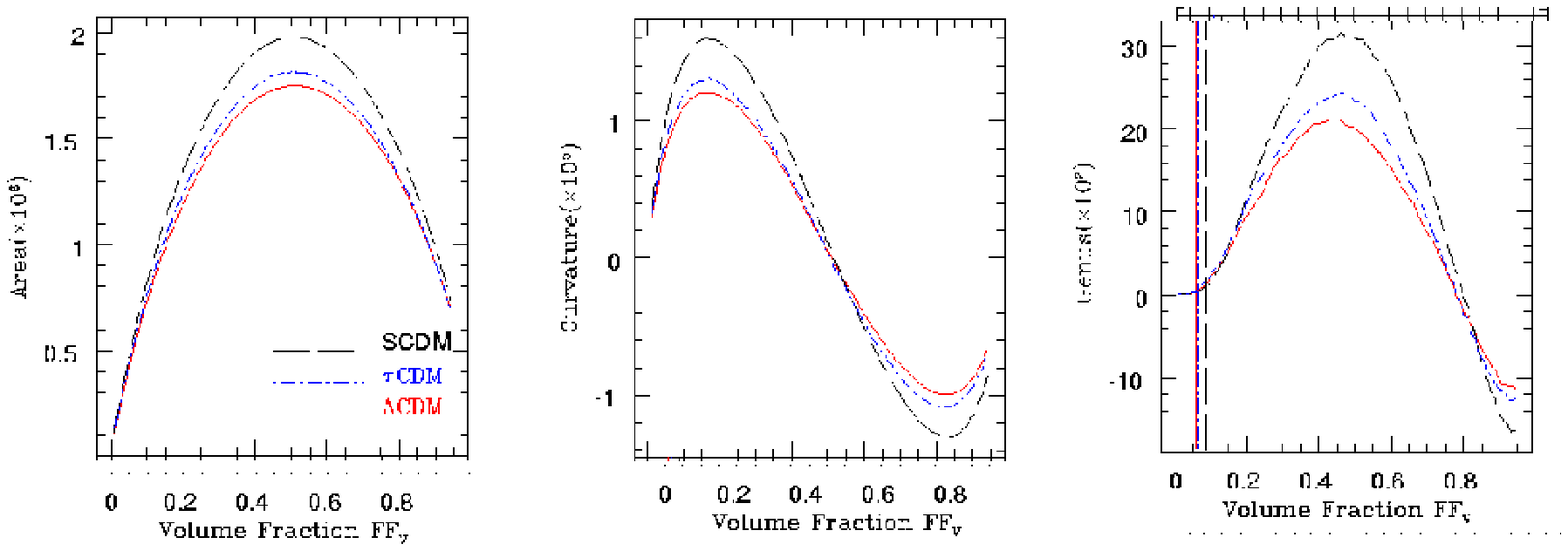}
    \caption{Three global MFs for $\La$CDM (solid lines), $\T$CDM (dot-dashed
      lines) and SCDM (dashed lines) are shown as functions of the
      overdense volume fraction FF$_{\rm V}$. The three models have
      appreciably different morphology and hence can be distinguished
      from one another on the basis of MFs. For futher discussion,
      please refer to the text. Figure courtesy of
      \cite{sh-s-s-sh03}.}
    \label{fig:3gmfs}
  \end{center}
\end{figure}
The three upper panels of Figure \ref{fig:3gmfs} show clustered
distribution of dark matter at $z=0$ in three rival cosmological
models $-$ SCDM, $\T$CDM and $\La$CDM respectively, simulated by Virgo
group \cite{jenkcdm}.  The three models are simulated such that the
observed abundance of rich clusters of galaxies can be reproduced in
{\em all} of them.  SURFGEN was employed to quantify the visually
apparent differences between these models. Global MFs were evaluated
at a set of density levels, and were studied with respect to FF$_{\rm
  V}$, the fractional {\em overdense} volume \cite{sh-s-s-sh03}. The
lower 3 panels of Figure \ref{fig:3gmfs} show variations in 3 global
MFs $-$ area, curvature and genus $-$ as density level is
progressively brought down. Notice that MFs for SCDM (dashed lines)
show the largest amplitudes, those for $\La$CDM (solid lines) show the
smallest amplitudes with $\T$CDM curves (dot-dashed lines) falling
between the two.  Since the number of rich clusters and their relative
spatial distribution in all the models is the same, this difference
can be attributed to difference in matter-distribution along sheets
and/or filaments connecting the clusters \cite{sh-s-s-sh03, sheth04}.
Global MFs are thus noted to be remarkable at discriminating between
models (also see \cite{schmal99}).

In Section \ref{sec:sec4} we noted that the distribution of galaxies
cannot be fully quantified using its 2$-$point correlation function
$\xi_{gg}(r)$ alone.  However, the galaxy distribution predicted by a
given cosmological model must {\em at least} reproduce the observed
$\xi_{gg}(r)$. Cole et al.(1998) generated {\em mock} catalogues of
galaxies in a variety of cosmogonies subject to the above constraint.
Thus, galaxies were selected from the simulated $z=0$
mass-distributions so that the observed $\xi_{gg}(r)$ is reproduced.
The catalogues were further used to construct {\em mock} 2dFGRS and
SDSS redshift surveys in such models \cite{cole2df98}. Since MFs
depend on the hierarchy of correlation functions, these can be relied
upon in distinguishing the {\em galaxy distributions} resulting from
different models.  Sheth (2004) made such a comparison between {\em
  mock} SDSS catalogues of galaxies from $\La$CDM and $\T$CDM models
by evaluating global MFs of density fields smoothed with a Gaussian
window 6 \h~Mpc wide.
\begin{figure}[h]
  \begin{center}
    \centering
    \includegraphics[scale=0.7,trim=5 150 5 150,clip]{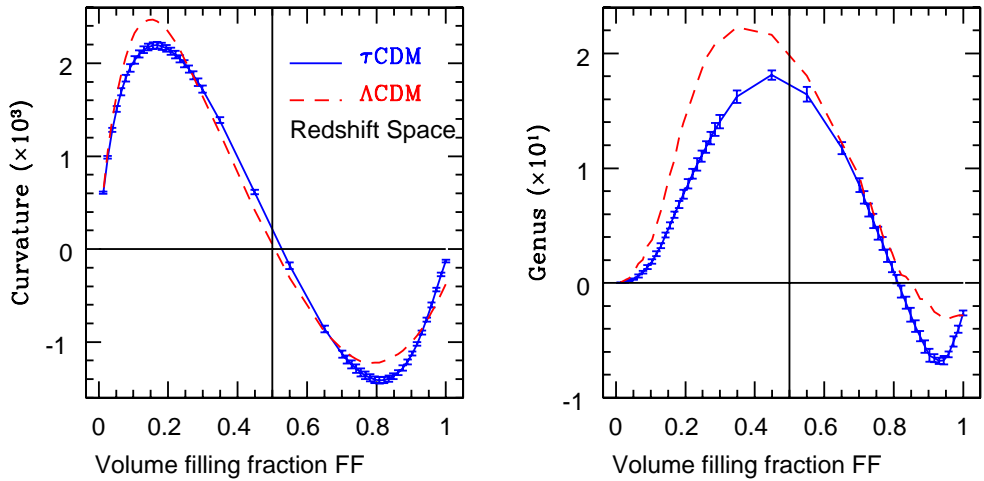}
    \caption{Two global MFs $-$ curvature
      and genus $-$ are evaluated at 50 density levels at a common set
      of volume filling fractions and are studied here with respect to
      FF$_{\rm V}$.  The values are normalised to the volume of [100
      $h^{-1}$Mpc]$^3$. ~The global MFs for $\T$CDM model are averaged
      over 10 realizations (solid lines) and the errorbars represent
      1$\sigma$ deviation.  Assuming the same level of accuracy for
      $\La$CDM, we may conclude that these MFs with volume
      parametrisation can indeed {\em clearly} distinguish $\T$CDM
      from $\La$CDM. Figure courtesy of
      \cite{jsh04}.}
  \label{fig:gmf_sdss}
  \end{center}
\end{figure}
Figure \ref{fig:gmf_sdss} shows the global mean curvature and genus
for the two models evaluated using SURFGEN. Note that the
MF-amplitudes in $\La$CDM are {\em larger} than those for $\T$CDM
model, and that the two models are successfully distinguished from
each other using MFs \cite{jsh04}, {\em even} in the case when the
distributions share the same $\xi_{gg}(r)$. (The $\T$CDM curves are
averaged over 10 realizations made available by \cite{cole2df98}.) MFs
can thus be reliably adopted to compare {\em mock} catalogues with the
observed LSS in redshift surveys.

\subsection{MFs and the scale-dependent bias}
We reviewed the status of our knowledge about bias in Section
\ref{sec:sec4}. We raised an important question as to whether the
clustering properties of dark matter and the {\em biased} galaxy
distribution could be different or otherwise. This question was
investigated by \cite{jsh04} by employing the mock SDSS catalogues of
galaxies in $\La$CDM and $\T$CDM models. The clustering properties
of galaxies were measured in terms of MFs.

As noted by \cite{melotopo90,springtop}, the amplitude of the
genus-curve drops as the N$-$body system develops phase correlations.
Given two density fields, the system with larger genus-amplitude shows
many more tunnels/voids which are, therefore, smaller in size. As time
progresses, the voids are expected to expand and merge, leading to a
drop in the genus-amplitude, while the phase correlations continue to
grow. With this simple model, one could correlate the amount of
clustering with the relative {\it smallness} of the amplitude of
genus, and therefore, of the area and the mean curvature. Going by
this reasoning, we may infer from Figure \ref{fig:gmf_sdss} that the
$\La$CDM galaxy-distribution is relatively {\it more} porous and would
show {\em less} coherence on large scales compared to that due to
$\T$CDM. As illustrated in Figure \ref{fig:3gmfs}(see
\cite{sh-s-s-sh03}), the dark matter distribution of $\T$CDM model due
to Virgo group shows considerably larger amplitudes for the MFs
compared to the $\La$CDM model, whereas, we find the reverse trend in
the MFs of {\em the galaxy distributions} due to the same two models.
Evidently biasing appears to be a source of this effect.

To establish this effect more firmly, the $\T$CDM and $\La$CDM dark
matter Virgo simulations were analysed by adopting the same resolution
($\ell_g$=3.5$h^{-1}$Mpc) and the smoothing scale
($\lambda_s$=6$h^{-1}$Mpc) as utilised in the analysis of mock
catalogues. 
So as not to introduce any bias due to redshift space distortions,
global MFs of the galaxy catalogues were computed in {\it real space}.
\begin{figure*}
  \centering \centerline{
    \includegraphics[scale=0.7,trim=20 140 5 140,clip]{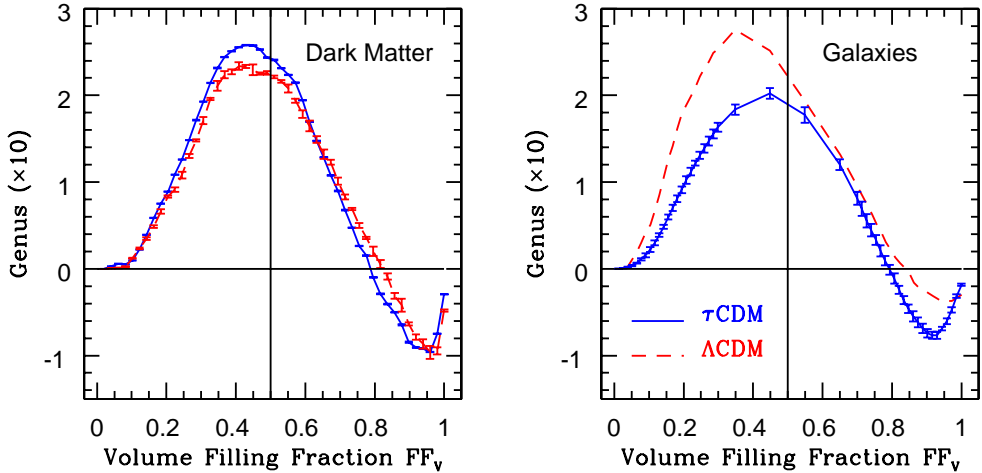}}
  \caption{{{\it Left panel:} Global genus curves of $\La$CDM and $\T$CDM {\em matter 
        distributions} are shown. The values refer to a volume of [100
      $h^{-1}$Mpc]$^{3}$. The 1$\sigma$ error-bars are due to 5
      realizations of both the models, each with 25\% of the total
      number of particles. Calculations are carried out on a grid of
      resolution 3.5 $h^{-1}$Mpc after smoothing the density fields
      with 6 $h^{-1}$Mpc Gaussian kernel. {\it Right panel:}The galaxy
      catalogues due to the two models are analysed in {\em real}
      space under identical conditions. This figure illustrates the
      phenomenon of {\em phase-reversal}. For more details, please
      refer to the text. Figure courtesy of
      \cite{jsh04}.}}
  \label{fig:realglo}
\end{figure*}
The effect of biasing is most dramatically seen in the global
genus-curve of the dark matter and galaxies (see Figure
\ref{fig:realglo}). We notice that the amplitudes ${\cal
  G}_{DM}^{\Lambda} < {\cal G}_{DM}^{\tau}$, whereas ${\cal
  G}_{G}^{\Lambda} > {\cal G}_{G}^{\tau}$. Here subscripts $DM$ and
$G$ stand for ``dark matter'' and ``galaxies'', respectively. We may
conclude from here that the matter distribution in $\La$CDM is less
porous than that in $\T$CDM, but this trend reverses as we investigate
biased distributions of galaxies due to the same two models: the
galaxy distribution due to $\La$CDM appears to be {\it more} porous
than that due to $\T$CDM. This could perhaps mean that a simple scheme
of scale-dependent biasing leads to different degree of
phase-correlations in the galaxy distribution as compared to the
underlying mass distribution. This can be dubbed as a phenomenon of
{\em phase-mismatch} between the two distributions $-$visible and dark
\cite{jsh04}.  This is supported by the fact that \cite{cole2df98}
require large anti-bias in high density regions of the $\La$CDM model.

To conclude, the study of the global MFs reveals that, local,
density-dependent bias could lead to an apparent {\em phase-mismatch}
or different phase-correlations among the dark matter and the
galaxies. It therefore becomes important to include realistic
treatment of bias before comparing theoretical predictions about LSS
with redshift surveys.
\begin{table*}
\centering
\caption{The ten most voluminous superclusters (determined at the 
  percolaton threshold) are listed with their associated Minkowski functionals
  (Volume, area, curvature and genus) and Shapefinders ${\cal T, B, L}$ in case
  of the $\La$CDM model simulated by Virgo group. The first row 
  describes the percolating supercluster and appears in boldface. It should 
  be noted that the interpretation of ${\cal L}$ as the `linear length' of a 
  supercluster can be misleading for the case of superclusters 
  having a large genus. In this case ${\cal L}\times (G+1)$ provides a more 
  realistic estimate of supercluster length since it allows for its numerous 
  twists and turns.}
\begin{center}
\begin{tabular}{llllllllllllllr} \hline 
\multicolumn{1}{l} {Model} & 
\multicolumn{1}{l} {Volume} & 
\multicolumn{1}{l} {Area} & 
\multicolumn{1}{l} {Curvature} & 
\multicolumn{1}{l} {Genus} & 
\multicolumn{3}{c} {Shapefinders ($h^{-1}$Mpc)} \\ 
 & $(h^{-1}Mpc)^3$ & $(h^{-1}Mpc)^2$ & $h^{-1}$Mpc & & ${\cal T}$ & ${\cal B}$ & ${\cal L}$\\
\hline
$\La$CDM & ${\bf 8.45\times10^4}$ &  ${\bf 4.5\times10^4}$ & ${\bf 6.16\times10^3}$ & {\bf ~~6} & {\bf 5.63} & {\bf 7.30} & {\bf 70.03} \\
$\delta=2.31$ & $3.21\times10^4$ & $1.7\times10^4$ & $2.33\times10^3$ & ~~~1 & 5.63 & 7.34 & 92.74\\
      & $2.22\times10^4$ & $1.22\times10^4$ & $1.76\times10^3$ & ~~~1 & 5.47 & 6.93 & 70.00 \\
      & $2.13\times10^4$ & $1.02\times10^4$ & $1.29\times10^3$ & ~~~0 & 6.27 & 7.90 & 102.7 \\
      & $1.21\times10^4$ & $6.35\times10^3$ & $8.78\times10^2$ & ~~~0 & 5.71 & 7.22 & 69.95 \\
      & $1.19\times10^4$ & $6.82\times10^3$ & $1.02\times10^3$ & ~~~1 & 5.25 & 6.70 & 40.53 \\
      & $1.08\times10^4$ & $6.12\times10^3$ & $9.16\times10^2$ & ~~~0 & 5.3 & 6.68 &  72.90 \\
      & $1.01\times10^4$ & $5.46\times10^3$ & $8.19\times10^2$ & ~~~0 & 5.57 & 6.65 & 65.24\\
      & $8.6\times10^3$  & $4.74\times10^3$ & $7.07\times10^2$ & ~~~0 & 5.44 & 7.00 &  56.33\\
      & $8.1\times10^3$ & $4.25\times10^3$ & $5.71\times10^2$  & ~~~1 & 5.72 & 7.44 &  22.74 \\
\hline
\label{tab:top10s123}
\end{tabular}
\end{center}
\end{table*}
We noted earlier that morphology of LSS appears web-like with frothy
voids separating sheet-like and/or filamentary superclusters. The rich
texture of LSS is a culmination of two interconnected factors: (1)
strong non-Gaussianity induced in cosmic density field due to
gravitational dynamics and (2) biased formation of galaxies relative
to the underlying mass. Superclusters are thus strongly non-Gaussian,
baryonic structures. Unlike galaxies and their clusters, superclusters
are as yet unrelaxed and dynamically evolving; their morphological
properties may be expected to be sensitive to the cosmological
parameters of the Universe.  It is hence important to quantify their
shapes and sizes and relate them with physically relevant other
quantities such as the mass enclosed or volume occupied by them.
Similarly, voids {\em too} can be studied, and the combined properties
of the supercluster-void network could provide us with yet another,
independent check on the cosmological parameters. Since modern
redshift surveys are deep and large enough, this may perhaps be the
most effective method to test theoretical predictions.

Recently, a comprehensive morphological study of the supercluster-void
network in $\La$CDM cosmogony was carried out by \cite{sh-sh-s04}.
Dark matter distribution at $z=0$ was smoothed with 5 \h~Mpc Gaussian
window to produce a cosmic density field sampled over a box of size
239.5 \h~Mpc. Superclusters and voids were defined as overdense and
underdense connected regions, respectively and their MFs and
Shapefinders were measured using SURFGEN. 

Superclusters in $\La$CDM model were found to percolate through the
Universe while occupying fractional overdense volume as small as 7\%
\cite{sh-sh-s04}. This points to the connectedness of the overdense
regions in the Universe. From their tendency to effectively percolate
the volume, it is apparent that superclusters must show strong
departure from sphericity. Voids, on the other hand, occupy as large a
fractional volume as 22\% before the underdense regions percolate
\cite{sh-sh-s04}.  Evidently, the morphology of voids would be
markedly different from superclusters. In shape, they are likely to be
more isotropic.

\begin{figure}[h]
  \centering \centerline{
    \includegraphics[width=5in]{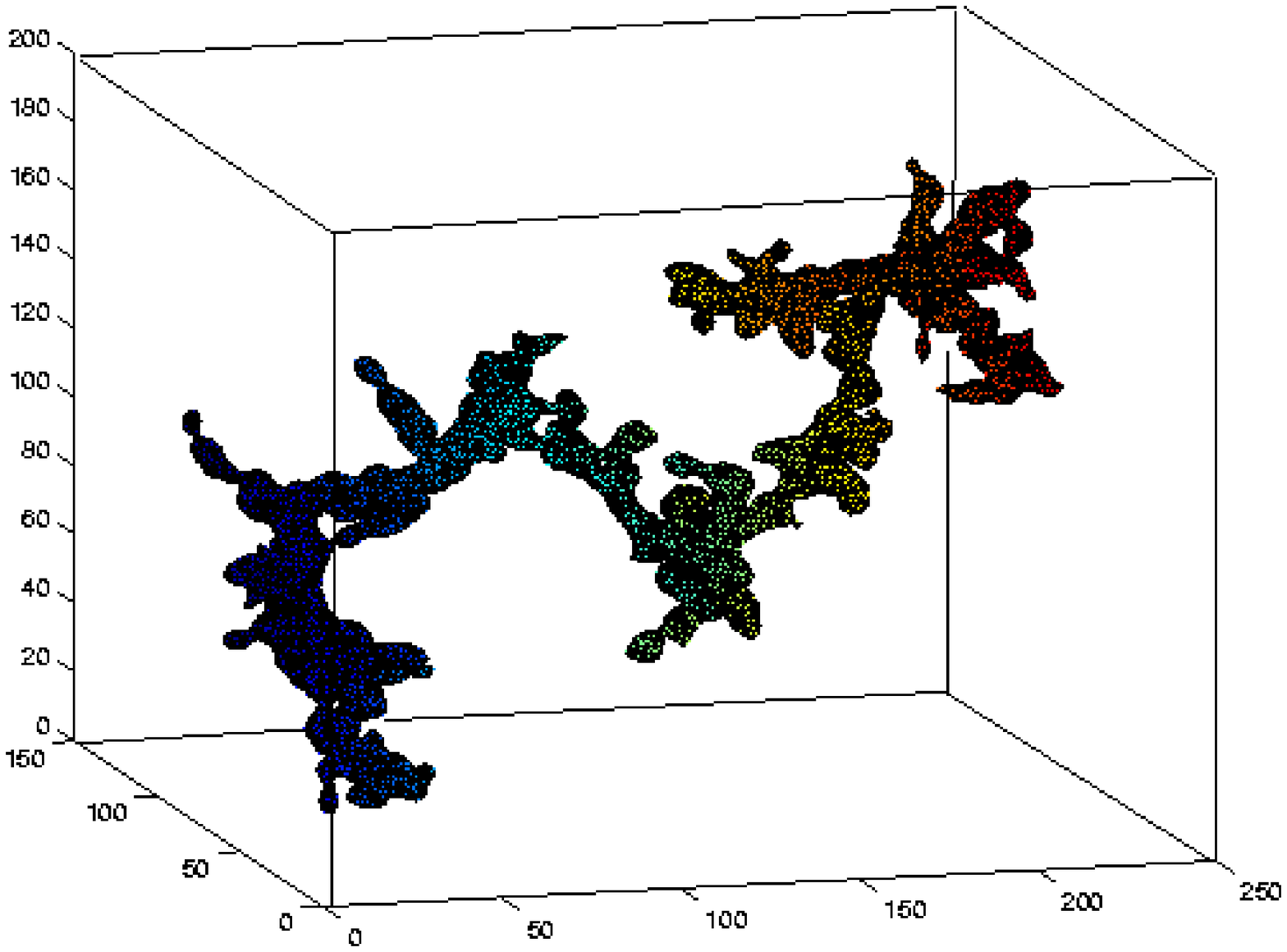} }
  \caption{{The largest (percolating) supercluster in $\La$CDM. 
    This cluster is selected at the density threshold which marks the
    onset of percolation ($\delta_{\rm perc} = 2.3$).  As demonstrated
    in the figure, the cluster at this threshold percolates through
    the entire length of the simulation box.  It is important to note
    that the percolating supercluster occupies only a small fraction
    of the total volume and its volume fraction (filling factor) is
    only $0.6\%$. Our percolating supercluster is a multiply connected
    and highly filamentary object. Its visual appearance is accurately
    reflected in the value of the Shapefinder diagnostic assigned to
    this supercluster: (\thi,\bre,\len)=(5.63, 7.30, 70.03) \h Mpc and
    $({\cal P,F},G) = (0.13, 0.81, 6)$. 
    Figure courtesy of \cite{sh-s-s-sh03}.}}
\label{fig:perclcdm}
\end{figure}

Shandarin, Sheth and Sahni (2004) further supported above inferences
by studying the sizes of superclusters and voids with respect to the
enclosed mass and occupied volume, respectively.
\begin{figure}[h]
  \begin{center}
    \centering
    \includegraphics[width=4.0in]{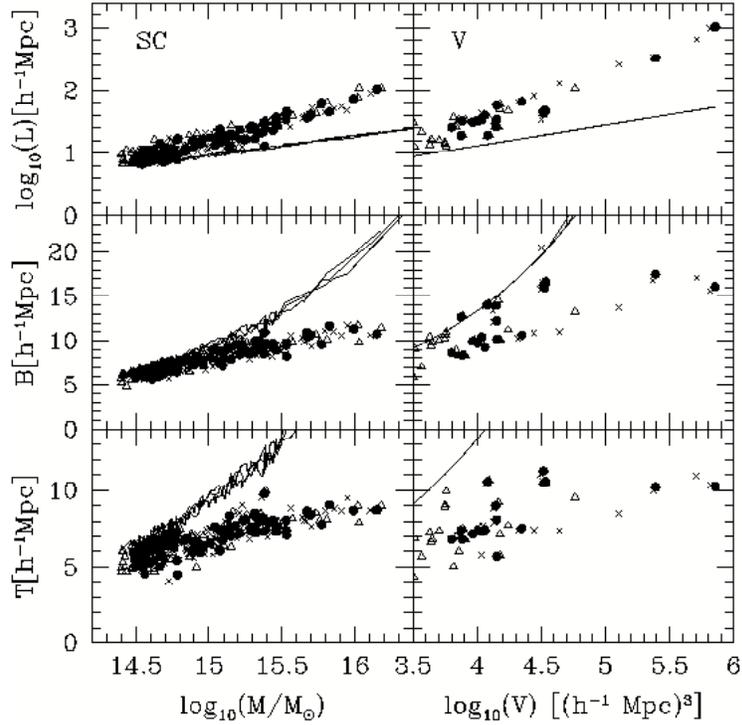}
    \caption{The length, breadth, and thickness versus mass for superclusters
      and versus volume for voids at percolation. Solid circles show
      the relation at percolation thresholds: $FF_C=0.07$ for
      superclusters and $FF_V=0.22$ for voids.  Crosses show the
      parameters before percolation ($FF_C=0.06$ for superclusters and
      $FF_V=0.21$ for voids) and empty triangles after percolation
      ($FF_C=0.08$ for superclusters and $FF_V=0.23$ for voids).
      Solid lines show the radius of the sphere having the same volume
      as the corresponding object.  Note the logarithmic scale used
      for the length. Three lines correspond to three different
      thresholds. Figure courtesy of
      \cite{sh-sh-s04}.}
    \label{fig:scvoid1p}
  \end{center}  
\end{figure}
Figure \ref{fig:scvoid1p} shows the results. Notice that all three
sizes of the structures show significant correlation with mass
enclosed (for superclusters) and volume occupied (for voids): the
larger the mass, the larger the size of SC and more the volume, the
larger is the size of the void, as expected. The solid lines show the
radius of a sphere having the same volume as a given object ($R=(3 V/4
\pi)^{1/3}$).  The thickness and breadth approximately double their
values and length grows by over an order of magnitude when the mass
increases from about $10^{14.5}\ M_{\odot}$ to $10^{16.5}\ M_{\odot}$.
Both the thickness and breadth are considerably smaller than the
radius $R$ of a sphere having similar volume for large superclusters
($M \ggeq 10^{15}\ M_{\odot}$).  On the other hand the length is
considerably greater than $R$.  This is a clear manifestation of the
anisotropy of large-scale superclusters (see Figure
\ref{fig:perclcdm}). Voids demonstrate similar correlations with their
volumes as shown in the right panels of Figure \ref{fig:scvoid1p}.
Voids, especially large ones, are also anisotropic, e.g., see
\cite{sh-sh-s04} for illustrations.  It was shown for example, that
{\em both} superclusters and voids show significantly larger
filamentarity in proportion to the mass enclosed and volume occupied,
respectively. Superclusters show negligible planarity whereas, voids
could be relatively more planar \cite{sh-sh-s04}. 

In their visual appearance, filaments stand out remarkably well in our
impression of LSS.  However, the occasional occurrence of {\em vast},
sheet-like superclusters such as the {\em SDSS Great Wall} 
\begin{figure}
  \begin{center}
    \centering
    \includegraphics[width=4.2in]{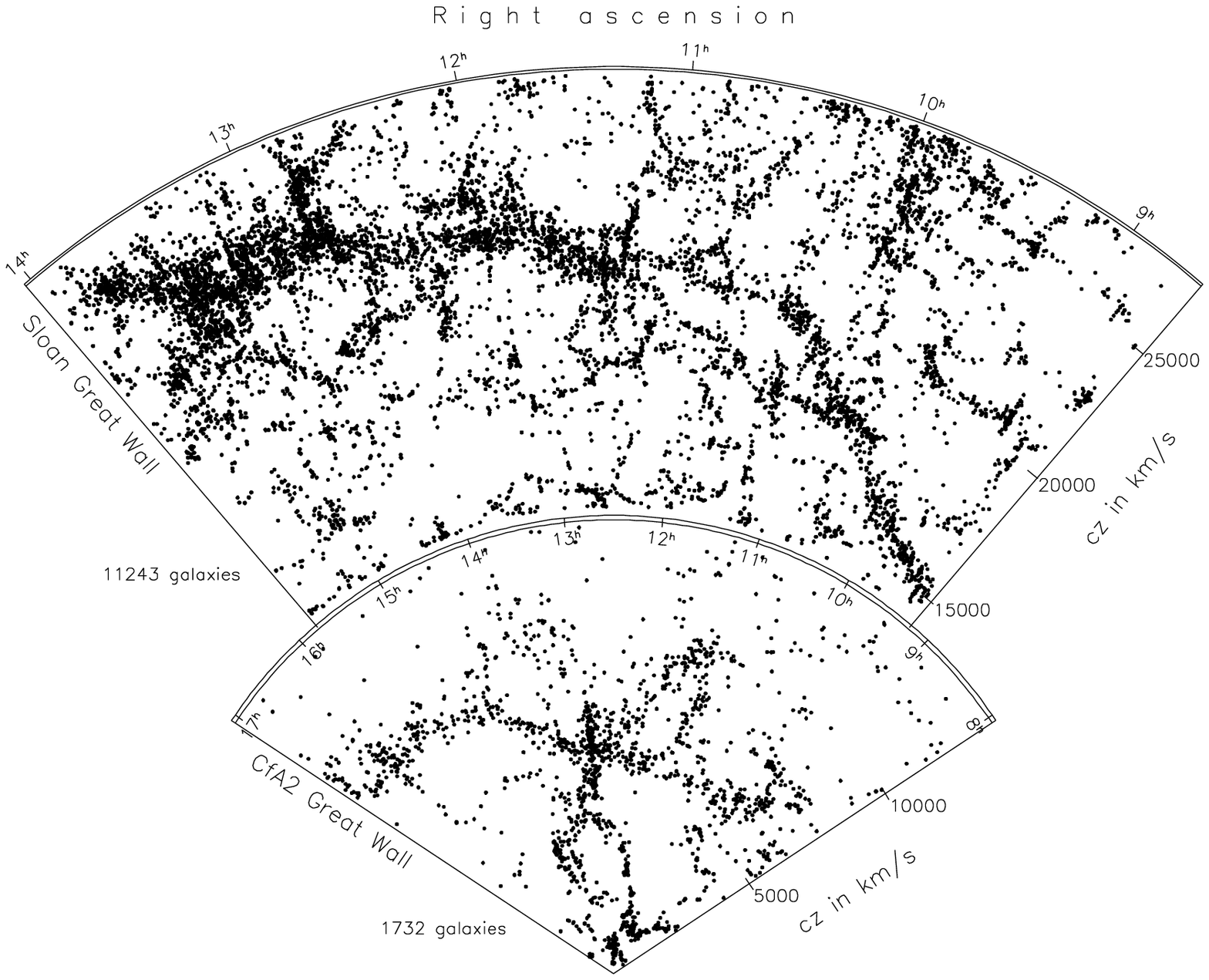}
    \caption{The {\em Sloan Great Wall} situated at $z\simeq0.08$ is shown 
      in the upper panel of the figure. For comparison, a relatively
      nearer structure of the CfA slice is also shown. The linear
      extent of the Sloan Great Wall is $\sim420 h^{-1}$Mpc, which is
      about twice larger than the CfA great wall. Figure courtesy of
      \cite{gott03}.}
    \label{fig:greatwalls}
  \end{center}
\end{figure}
(which is almost twice as large as the CfA Great Wall
\cite{gott03,tegmark04a}; see Figure \ref{fig:greatwalls}.) raises
questions as to how significant are sheets with respect to filaments
on large scales in the {\em actual} Universe. In future, if the
sheet-like superclusters in redshift surveys are confirmed and their
counterpart structures {\em not} found in N$-$body simulations, it may
present a clear dichotomy between theory and observations.

To investigate this question in the context of simulations,
\cite{sh-s-s-sh03} studied the correlation in the Shapefinders of
$\La$CDM superclusters identified at percolation.
\begin{figure}[h]
  \begin{center}
  \centering \centerline{ \includegraphics[scale=0.9,trim=20 140 5
    140,clip]{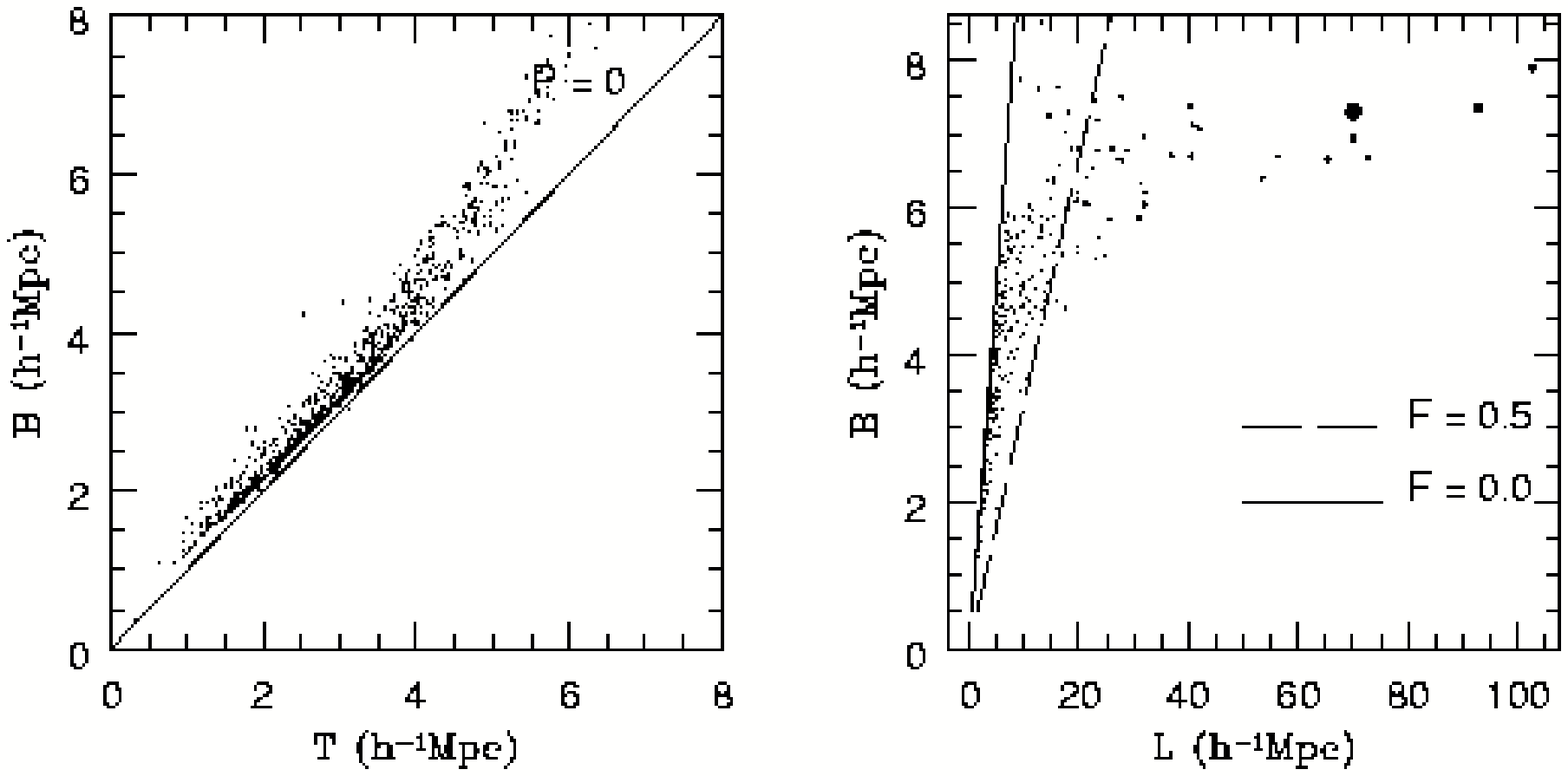} }
\caption{{Scatter plot for the
    pair of Shapefinders \thi~, \bre~ (left panel) and \bre~, \len~
    (right panel) defining the morphology of clusters/superclusters in
    the $\La$CDM model.  The strong correlation between \thi~ and
    \bre~ in the left panel near the line ${\cal P} = 0$ indicates
    that two of the three dimensions defining a cluster are equal and
    of the same order as the correlation length ($\simeq$ few Mpc.).
    Judging from the left panel we find that clusters/superclusters in
    $\La$CDM are either quasi-spherical or filamentary (since both
    satisfy \thi$\simeq$\bre $\Rightarrow {\cal P} \simeq 0$). The
    degeneracy between spheres and filaments is lifted by the right
    panel which is a mass-weighted scatter plot for the Shapefinders
    \bre,\len.  Each dot in this panel refers to a cluster and its
    area is proportional to the fraction of mass in that cluster.  The
    concentration of points near the line ${\cal F} = 0$ (\bre = \len)
    reflects the fact that a large number of smaller clusters are
    quasi-spherical.  The more massive structures, on the other hand,
    tend to be filamentary and the largest and most massive
    supercluster has ${\cal F} = 0.81$.  All objects are determined at
    the percolation threshold. Figure courtesy of
    \cite{sh-s-s-sh03}.}}
\label{fig:corr_5fs123}
\end{center}
\end{figure}
Figure \ref{fig:corr_5fs123} is a scatter plot of Shapefinders \thi,
\bre, \len~ for clusters in $\La$CDM defined at the percolation
threshold.  The strong correlation between \thi~ and \bre~ in the left
panel indicates that two (of three) dimensions defining any given
cluster assume similar values and are of the same order as the
correlation length.  (\thi$\simeq$\bre$\simeq$5\h Mpc for the largest
superclusters.)  The clustering of objects near
\thi$\simeq$\bre(${\cal P} \simeq 0$) in this panel suggests that the
superclusters are either quasi-spherical or filamentary.  The
superclusters in simulations thus {\em do not} appear to be planar.
The scatter plot between \bre~ \& \len~ in the right panel of Figure
\ref{fig:corr_5fs123} breaks the degeneracy between spheres and
filaments. The mass-dependence of morphology is highlighted in this
panel in which larger dots denote more massive objects. This figure
clearly reveals that more massive clusters/superclusters are, as a
rule, also more filamentary, while smaller, less massive objects, are
more nearly spherical \cite{sh-s-s-sh03}.

For Shapefinder-like statistics, it is important that the measured
properties of an object also conform with its visual appearance. 
Figure \ref{fig:perclcdm} shows the largest percolating supercluster
in the $\La$CDM cosmogony for which ${\cal F}=0.81$
\cite{sh-s-s-sh03}.  Note that the supercluster is indeed highly
filamentary in agreement with the measured, high value of ${\cal F}$.

Earlier we noted results from MF-based study of mock SDSS catalogues
due to $\La$CDM and $\T$CDM models. Global MFs were shown to
successfully discriminate between the two models. Further, we noted
the phenomenon of {\em phase-mismatch} between dark matter and
galaxies due to biasing.  In this subsection, we noted that
superclusters in a typical CDM-cosmogony are generically filamentary.
Could we use the {\em length} of the filamentary superclusters as
discriminatory statistic between models? In other words, are the
morphological properties of superclusters {\em indeed} sensitive to
the underlying cosmic parameters? The 10 most voluminous superclusters
in dark matter Virgo simulations of $\La$CDM, SCDM and $\T$CDM models
were shown to have considerably different morphological properties.
In fact, the percolating $\La$CDM supercluster (shown in Figure
\ref{fig:perclcdm}) was found to be more filamentary and
topologically more simpler ($G=6$) than its counterpart structures in
SCDM and $\T$CDM ($G\simeq20$).

\begin{figure*}
  \centerline
  \centering
  {\includegraphics[scale=0.7,trim=20 140 5 140,clip]{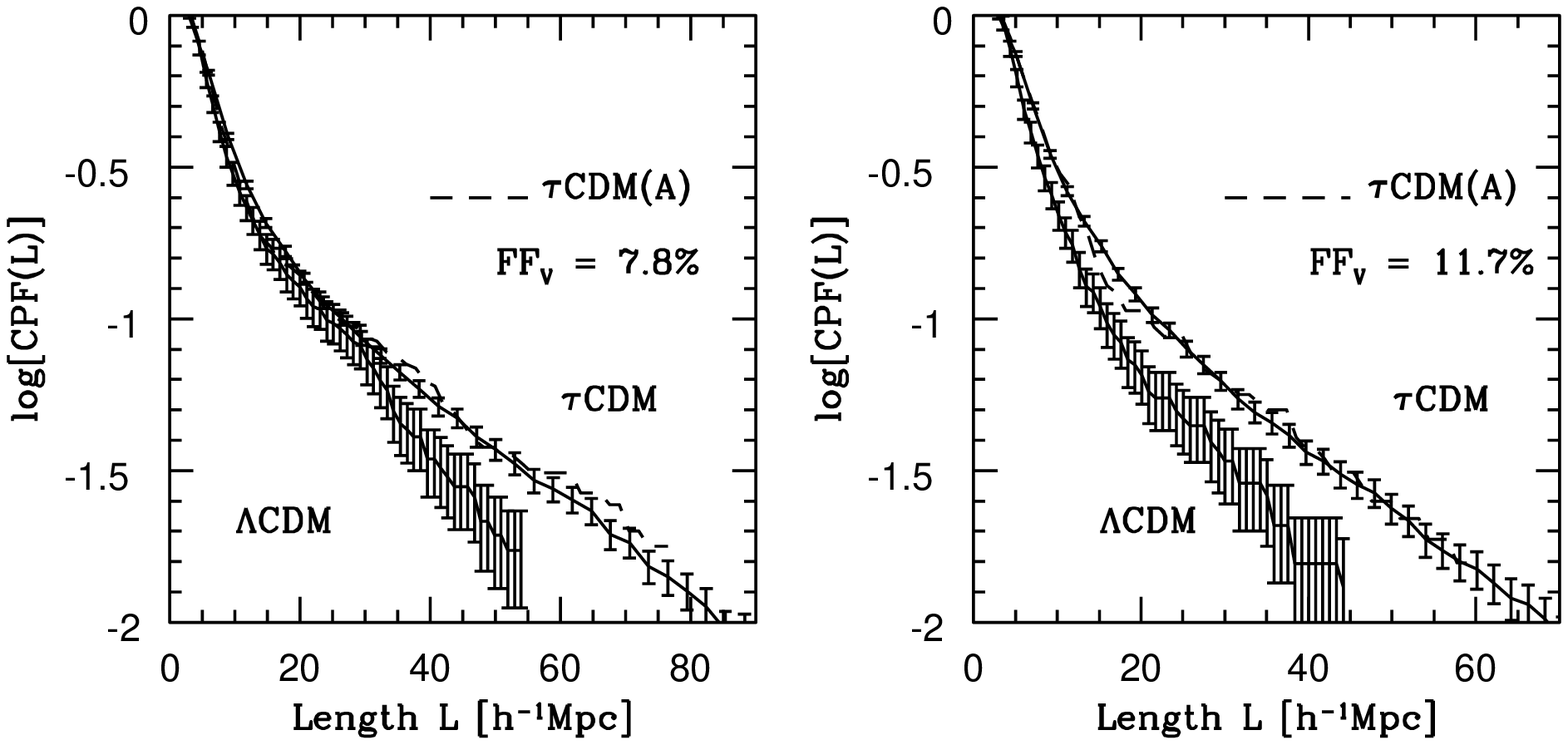}
    }
  \caption{{Shown here are the cumulative probability functions of
      length ${\cal L}$ for superclusters of {\em galaxies} in $\T$CDM
      and $\La$CDM at two thresholds of density corresponding to
      FF$_{\rm V}$=7.8 per cent (just before the onset of percolation)
      and FF$_{\rm V}$=11.4 per cent (after the onset of percolation).
      We find the CPFs of the two models to be distinctly different at
      longer length-scales.  The dashed line refers to the CPFs due to
      the first realization of $\T$CDM which shares the initial set of
      random numbers with $\La$CDM. The largest superclusters of
      $\T$CDM are systematically larger than those due to $\La$CDM.
      Figure courtesy of \cite{jsh04}.}}
  \label{fig:cpflen}
\end{figure*}

To answer the same question in case of galaxy distributions,
\cite{sheth04} studied the abundance of superclusters larger than
certain length ${\cal L}$ from within a given large ensemble of
superclusters selected from mock SDSS catalogues of $\La$CDM and
$\T$CDM models. This quantity is referred to as Cummulative
Probability Function (CPF). Figure \ref{fig:cpflen} shows the results,
where the CPF(${\cal L}$) have been evaluated at FF$_{\rm V}$=7.8 per
cent (at the onset of percolation) and at FF$_{\rm V}$=11.7 per cent
(after the percolation). Note that at the onset of percolation, the
length of the longest $\T$CDM superclusters could be as large as 90
$h^{-1}$Mpc, whereas their $\La$CDM counterpart structures, which
exhibit same degree of statistical significance, are relatively
shorter with ${\cal L}_{\rm max} = $55 $h^{-1}$Mpc. We conclude that
the $\T$CDM superclusters tend to be statistically longer than their
$\La$CDM counterpart structures. The large-scale coherence in the
superclusters is attributed to the phase-correlations in the density
field; the higher the degree of phase-correlations, the larger the
length-scale of coherence. Based on this, and the results reported
earlier, we would anticipate the $\T$CDM superclusters to be {\it
  longer than} those in $\La$CDM. As we can see, the result reported
here agrees well with this anticipation. This is a {\em considerable}
success for the ansatz of the Shapefinder quantifying {\em length},
for as we noted above, it helps us capture the relative effect of
phase-correlations among rival models of structure formation.

\subsection{Applications to Redshift Surveys}

We end this section by providing a brief summary of the results which
have emerged out of the analysis of the observed LSS. We further
outline the scope of these methods in light of the newly available
large datasets. 

The methods of topological analysis of LSS have been well developed
since late 1980s \cite{melotopo90}. Hence, most of the investigations
complementing the standard correlation function based approach are in
the wake of studying the connectivity and topology of LSS.  Almost all
the topological analyses are carried out to test whether the
galaxy-distribution is consistent with the underlying Gaussian,
random-phase primordial density field. Topology of LSS revealed both
in LCRS data \cite{colley97} and SDSS data \cite{hoyleetal02} has been
found to be consistent with Gaussian, random-phase hypothesis for the
primordial density fluctuations. SDSS Early Data Release has been
further found to agree well with the predictions of $\La$CDM
concordance cosmology model \cite{hikage02}. Springel et al.(1998)
reached similar conclusions based on their analysis of the 1.2 Jy
redshift survey \cite{springtop}. Connectivity of LSS of LCRS has been
studied by \cite{shyes98}. These authors detected a network of
filamentary superclusters in LCRS slices, which they found to be
consistent with the gravitational instability paradigm of structure
formation.

Recently there has been an upthrust of interest in complementing
earlier topological studies with geometry of LSS. Probably the first
morphological analysis of LSS was done using 1.2 Jy redshift survey
data by \cite{sathya98}. MFs for SDSS early data release (EDR) have
been evaluated by \cite{hikage03a} who find the data to be consistent
with the clustering predictions of the $\La$CDM cosmogony \cite{fn5}.
Basilakos (2003) has measured shapes of superclusters in SDSS(EDR) and
finds them to be predominantly filamentary. Due to their large volume,
SDSS and 2dFGRS can be fruitfully employed to test predictions for MFs
in a weakly nonlinear regime. An interesting application has been
illustrated by \cite{colley00}, who use mock SDSS catalogues (inspired
by $\La$CDM cosmology) for the purpose.

Morphology of LSS in LCRS has been studied by \cite{bharad} who detect
significant filamentarity in LCRS slices compared to an equivalent
random distribution. The LSS-data from recent redshift surveys such as
SDSS and 2dFGRS yet await an integrated morphological analysis and a
detailed comparison with theoretical predictions. Meanwhile,
encouraging efforts have been recently made to constrain the scale of
homogeneity of the LSS by investigating the largest {\em length-scale}
of filaments in LCRS slices \cite{bharad04a}. Robust 2-dimensional
morphological statistic such as the Shapefinder ${\cal F}$ have been
used for the purpose. The scale of homogeneity is found to be
$\simeq$70$-$80 \h~Mpc \cite{bharad04a}. $\La$CDM {\em mock} LCRS
catalogues have been further confronted with the LCRS slices using the
same techniques. A uniform, moderate large scale bias of 1.15 seems
to give good agreement between the model and the LCRS slices.

\markright{Bibliography}


\begin{thebibliography}{}
\bibitem{fn1} Morphology is a combined study of sizes and shapes of
  structures.  Connectivity, geometry and the overall shape of the
    structures form the subject matter of such a study.
\bibitem{fn2} Earlier these were already estimated by \cite{bauef93}
  and \cite{baugh96}, respectively by working with the 2-dimensional
    galaxy-counts recorded in the APM catalogue. Also see \cite{mad96}.
\bibitem{fn6}The two point correlation function completely describes
the properties of a gaussian random field.  Although the present
distribution of both baryons and dark matter arose from an initially
`featureless' Gaussian distribution, the process of gravitational
clustering gives rise to significantly non-Gaussian features
(superclusters, voids, etc.)  to describe which one needs to go
beyond the simple two point correlation function to higher order
correlation function's or, alternatively, by means of geometrical
diagnostics such as the genus, percolation theory and more generally
the Minkowski functionals which will be described below.
\bibitem{fn7} $FF$ is the cumulative probability distribution
  function: $FF(\delta_T) = P(\delta > \delta_T)$.
\bibitem{fn5}It is worth mentioning here that these authors do not
  account for nontrivial effects of bias in their treatment.
\end{thebibliography}
\end{document}